\documentclass[twocolumn]{svjour3}  
\smartqed  

\usepackage[dvipdfmx]{color}
\usepackage{graphicx}
\usepackage[T1]{fontenc}
\usepackage{mathptmx}  
\usepackage[scaled]{helvet}  
\usepackage{courier}
\usepackage[subrefformat=parens]{subcaption}

\makeatletter
\let\cl@chapter\undefined
\makeatletter
\usepackage{amsmath,amssymb}   
\usepackage{mathtools} 
\usepackage{cleveref}  
\Crefname{equation}{Eq.}{Eqs.}%
\Crefname{figure}{Fig.}{Figs.}%
\usepackage{amsbsy}
\usepackage{bm}
\usepackage{algorithmic}
\usepackage{algorithm}

\usepackage{url}

\journalname{Journal of Marine Science and Technology}
\begin{document}

\title{Perspective on the Marine Simulator for Autonomous Vessel Development
}

\author{Ryouhei Sawada \and Yoshiki Miyauchi \and Suisei Wada \and Takuya Tanigushi \and Satoru Hamada \and Hiroaki Koike \and Kouki Wakita \and Atsuo Maki}

\institute{Ryouhei Sawada \at
           National Maritime Research Institute, 6-38-1, Shinkawa, Mitaka, Tokyo, Japan\\
           \email{sawada-r@m.mpat.go.jp}
           \and
            Yoshiki Miyauchi \and Suisei Wada \and Takuya Tanigushi \and Satoru Hamada \and Hiroaki Koike \and Kouki Wakita \and Atsuo Maki \at
              Osaka University, 2-1 Yamadaoka, Suita, Osaka, Japan \\
              \email{maki@naoe.eng.osaka-u.ac.jp} 
}

\date{Received: date / Accepted: date}

\maketitle

\begin{abstract}
There is a growing demand for simulators for the research and development of maritime autonomous surface ships (MASS) and the approval of autonomous navigation algorithms.
Simulators are used for purposes such as evaluation and training and are taken on various configurations accordingly. 
The ship maneuvering mathematical model used in such a simulator is an important element that characterizes the simulator.
In this paper, we discuss the dynamic model of the hull and its position in the simulator that will be required for MASSs in the future. 
It also discusses guidelines for selecting an appropriate model, which has not been discussed extensively in previous studies.
Finally, we discuss the functional requirements that simulators should have.

\end{abstract}

\section{Introduction}\label{sec:introdction}

\begin{figure}[tb]
    \centering
    \includegraphics[width=1.0\hsize]{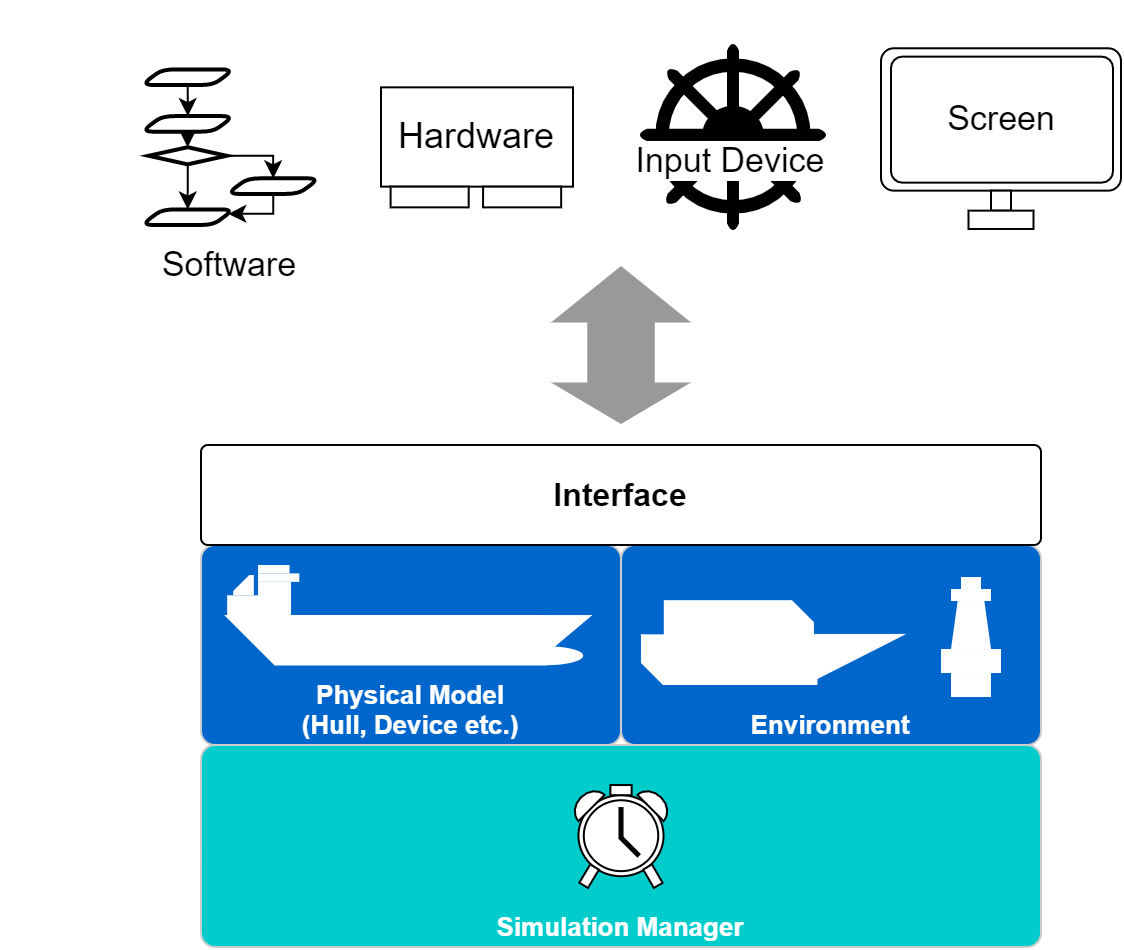}
    \caption{Nominal structure of the ship maneuvering simulator}
    \label{fig:simulator_general_structure}
\end{figure}

In the research and development of maritime autonomous surface ships (MASSs), there is a growing demand for marine simulators in research institutions and manufacturers. As described later, simulators are used for purposes such as evaluation and training of operators, taking various configurations depending on the intended use. 
Generally, they have the physical models of the ship motion, environment, and navigational instruments as illustrated in \cref{fig:simulator_general_structure}.

Among these functions, the dynamic model of the hull (or maneuvering mathematical model, control model, equation of maneuvering motion) can be considered a crucial element characterizing the simulator.
However, in the development of autonomous ships, the discussion of model selection has become complex, involving not only simulations with the dynamic model of the hull alone but also simulations of dynamic systems incorporating the control system.
Thus, this paper aims to describe the role of the dynamic model of a ship, particularly among the functions of the simulator, in the context of what will be required for autonomous ships.

In recent years, the role of simulators in the development of autonomous ships has increased. 
Ship classes have published guidelines for autonomous ships including simulation\cite{ABS_guide,NK_guide,BV_guide,dnv_guide}, whereas the requirements of simulators to evaluate ship motion and autonomous navigation systems have not been adequately discussed.
Meanwhile, DNV-GL's standard for maritime simulator systems \cite{dnv_sim} delineates requirements for the performance of maritime simulator systems required for simulator-based training or demonstrations mandated by the International Convention on Standards of Training, Certification and Watchkeeping for Seafarers (STCW). This standard includes the requirements for hydrodynamic ship models within a certified simulator. For instance, this standard gives the requirements for the speed data of each hydrodynamic ship, however, it doesn’t include low-speed ship speed data for berthing operations. The standard also requires maritime simulators to possess the capability to simulate physical and behavioral realism, without, however, specifying a reference model or implementation.

It extends beyond the conventional use of maritime training simulators, with the development of dedicated simulators serving as environments for developing autonomous ship systems. For instance, platforms like the Open Simulation Platform (OSP) \cite{osp_PEDERSEN2020104799} and the Fast Time Ship Simulator (FTSS) developed by the National Maritime Research Institute \cite{Minami_2022} have emerged.
In the development of autonomous ships, the purposes for using simulators may include the following:
\begin{description}
    \item[$\bullet$] Confirmation of the hydrodynamic performance of a ship maneuvering motion.
    \item[$\bullet$] Tuning of parameters such as the gain of model-free control.
    \item[$\bullet$] Training crews.
    \item[$\bullet$] Learning of image recognition or control.
    \item[$\bullet$] Evaluation of controller 
    \item[$\bullet$] Creation of reference routes for tracking maneuvers during berthing/unberthing.
    \item[$\bullet$] Assessment of System Functional Safety (Integration Testing) 
\end{description}
To address these diversified purposes of simulator use, this study discusses ship dynamic models suitable for simulators in the development of autonomous vessels.

In particular, the selection of the ship's dynamic model is a challenge within this context.
The functional requirements of the model change as the operational scenarios vary, for example, when simulating different operational conditions such as berthing and ocean navigation.
Furthermore, the desired dynamic model will depend on which functions of the system are to be evaluated by the simulation.
However, there is little guidance or discussion as to what model is appropriate to choose in the context of developing such functions.

As research and development on autonomous ships becomes more active, the opportunities for selecting and creating dynamic models are increasing.
In this paper, the authors argue that the model used in the simulator should have requirements that satisfy the objective of simulation and that the model is only required to meet those requirements.
For instance, in the case of collision avoidance, a relatively simple model may be sufficient, while in scenarios of port entry and departure, a model is required to consider wind disturbances, forward and backward motions at low speed, thrust near obstacles, and shallow/sidewall effects.
In developing control systems, developers are interested in minimizing the modeling errors and compensating for them by the controller.
A review of international literature reveals a trend toward simpler approaches, different from the MMG (Mathematical Maneuvering Group) model. For example, considering the research group led by Fossen at NTNU, the models consistently used there are often more straightforward compared to MMG.

The construction of a dynamic model requires activities such as towing tank tests, numerical simulations, and free-run tests using a full-scale ship or a model ship, all of which require time and resources.
Developers generally expect high functionality for the model, but as the complexity of the model increases, so does the cost of model construction.
In other words, there is generally a trade-off between the two. Therefore, defining requirements based on the purpose of the simulation at the outset is believed to optimize the cost of model construction.

In this study, We will summarize models in real-world use cases from previously published papers, focusing primarily on controls and simulators.
We aim to provide solutions and ideas that can assist engineers and researchers engaged in or aspiring to conduct modeling, aiding them in the selection of models.


The structure of this paper is as follows. Firstly, in \cref{sec:objectives_model}, we organize the objectives of creating dynamic models. We briefly explain the history of dynamic models in the maritime field and their applications proposed in previous studies. 
In \cref{sec:history_categorization}, we provide an overview of the classification and history of dynamic models. \Cref{sec:modeling_cost} describes the trade-off between the complexity of a dynamic model and its construction costs, using specific model calculations. In \cref{sec:functional_requirement}, we discuss examples of functional and non-functional requirements for simulating autonomous ships. \Cref{sec:ODE_solve} addresses tips when solving differential equations. In \cref{sec:discussion}, we conduct a discussion based on these results, and the conclusion is presented in \cref{sec:conclusion}.

Initial results from the investigation described in this study were originally developed by Miyauchi et al. \cite{Miyauchi2023Sim_JASNAOE}. In this paper, the results are presented more extensively, with more details and some revisions.





\section{Notation and coordinate systems}\label{sec:notation}

In the following, the set of real numbers is denoted by $\mathbb{R}$, and the $n$-dimensional Euclidean space is represented as $\mathbb{R}^n$. The prime symbol $'$ indicates a non-dimensional value, while the overdot $\dot{}$ represents a time derivative.

\begin{figure}
    \centering
    \includegraphics[width=1.0\hsize]{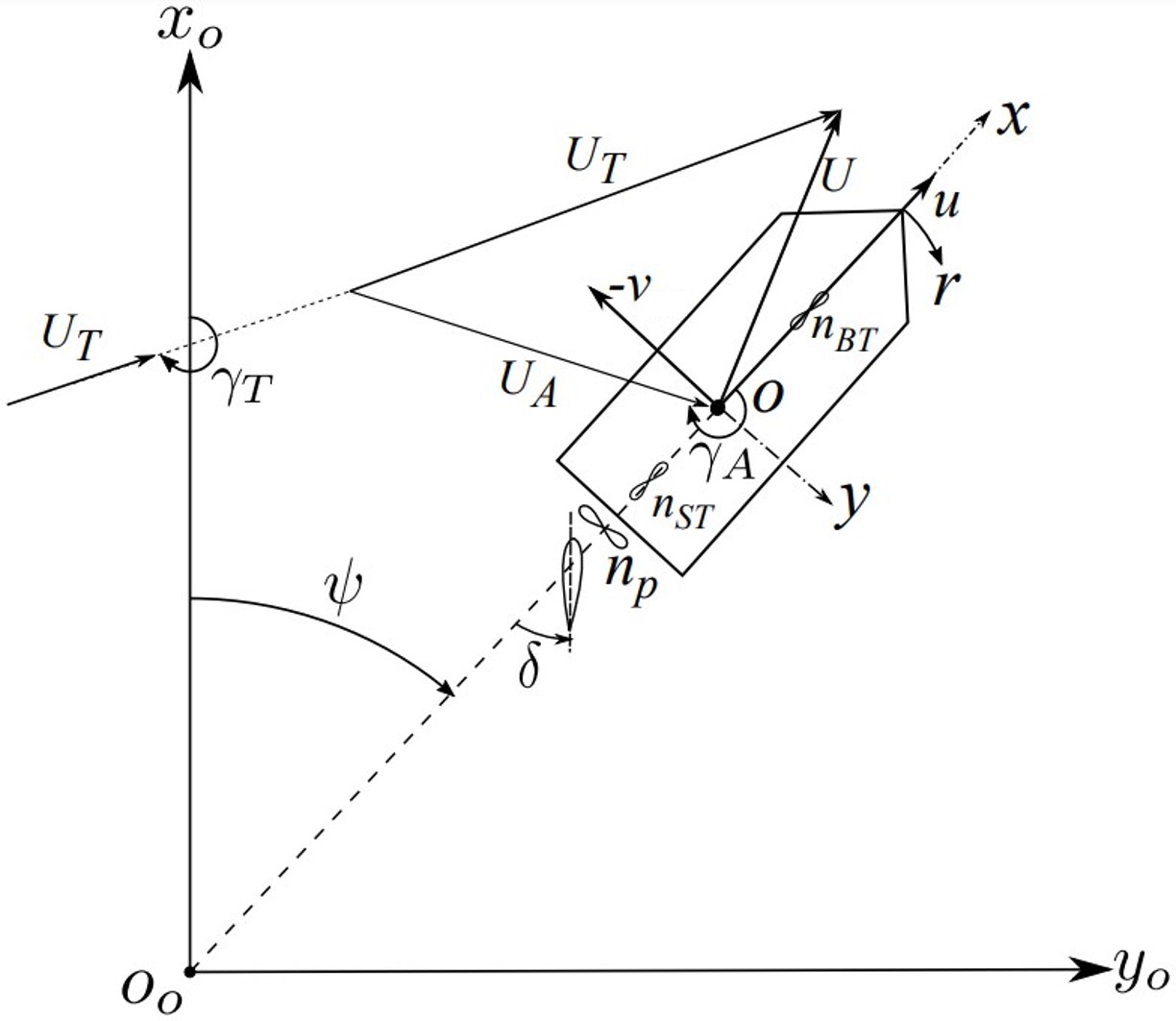}
    \caption{Coordinate System. The coordinate system consist by space-fixed system $O_{0}-x_{0}y_{0}$ and ship-fixed system $O-xy$. Notations $u,v_{m}$ are the velocity of $O-xy$ system;  $\delta,~n_{\mathrm{p}},~n_{\mathrm{BT}},~n_{\mathrm{ST}}$ are the rudder angle, the revolution of the propeller, the bow thruster and the stern thruster; $\gamma_{T},~U_{T}$ are true wind direction and speed; $\gamma_{A},~U_{A}$ are apparent wind direction and speed.}
    \label{fig:coordinate}
\end{figure}

With a few exceptions, this paper does not deal with tidal currents.
The approach for currents can be found in the references \cite{Fujino1983_En,Shouji1989_En}.
This paper focuses on surface ships, and its coordinate system is shown in\cref{fig:coordinate}.
In this paper, forces and moments shall be non-dimensionalized as follows, unless otherwise noted:
\begin{equation}
    \left\{
    \begin{aligned}
        \text{Force}&:\,\, \frac{1}{2}\rho U^2 L d\\
        \text{Moment}&:\,\, \frac{1}{2}\rho U^2 L^2 d\\
    \end{aligned}
    \right.
    \label{eq:nondimentionalization}
\end{equation}
where $\rho$ is the density of fluid, $L$ is the length of a ship and $d$ is the draft.
The resultant velocity $U$ is defined as follows:
\begin{equation}
    U \equiv \sqrt{u^2+v^2}.
\end{equation}
The drift angle $\beta$ defined as follows, except when $U=0$:
\begin{equation}
    \beta \equiv -\sin^{-1} \left(\frac{v}{U}\right)
\end{equation}

\section{Objective of making a dynamic model}\label{sec:objectives_model}

In the beginning, the authors classify dynamic models based on the purpose of the models. For instance, the ITTC recommended procedure \cite{ITTC2021manuvalidation} distinguishes maneuvering simulation models into two categories: 1. Models for predicting ship maneuvering performance and evaluating ship design by estimating standard maneuvers defined by IMO, and 2. Models for use in simulators, intended for real-time man-in-the-loop simulations, typically used for training ship crews, etc. Additionally, the literature \cite{fossen2021handbook} classifies dynamic models into simulation models, control algorithm models, and state estimator design models, depending on their applications.

In this study, we classify dynamic models into four categories: 
\begin{enumerate}
    \item Models for evaluating maneuvering characteristics during ship hull design,
    \item Models for ship-handling simulator,
    \item Model for algorithm development and performance validation of autonomous ships, and
    \item Models for embedding into model-based control laws or observers. 
\end{enumerate}
Below are the definitions for each category:

\paragraph{Models for evaluating maneuvering characteristics during ship hull design}
The maneuvering performance of a ship is a crucial aspect related to safety, and it is essential to understand specified performance criteria, such as turning and course-keeping capabilities mentioned in the IMO Steering Criteria MSC137(76), before construction. Simulation is one of the representative methods for estimating these performances, and dynamic models are used in this context.
Dynamic models for this purpose are desirable to be capable of considering the scale effects of full-scale and model-scale ships, and they should be easy to handle when making significant changes such as altering the ship's main particulars or rudder configuration.
In Japan, modular-type fluid dynamics models, such as the MMG model, which can account for the scale effects, are widely used for this purpose.



\paragraph{Models for ship-handling simulator}
A ship-handling simulator is a real-time simulator used for purposes such as crew training and reproducing accident scenarios. 
It replicates the functionality of a ship's bridge by connecting operational inputs to physical controls such as engine telegraph levers and steering wheels, allowing individuals to manipulate simulator input values. 
The ship-handling simulators do not need to accurately reproduce maneuvering characteristics; qualitative characteristics such as course instability need only be represented.
However, a model capable of real-time calculations with relatively short time steps is necessary. 
Additionally, if training and evaluation can be conducted in challenging environmental conditions (e.g., wind and waves) that are difficult to replicate on a real ship, the advantages of the simulator can be further maximized. 
Thus, it is desirable for the model to be in a format that can flexibly incorporate a variety of environmental disturbance factors. 
Dynamic models used in systems for Man-in-the-loop simulations for crew training have been discussed, for example, in \cite{Hara1981_En,Oliveira2022}.

\paragraph{Model for algorithm development and performance validation of autonomous ships}
The purpose of simulation as a development environment is to evaluate the performance and safety of autonomous ship control algorithms before assessing them on full-scale vessels.
It is crucial to verify how the algorithm behaves in a real system to evaluate its performance, and the accuracy of the simulation model is essential.
A Simulator also requires real-time capabilities since it is typically used for developing systems that operate in real-time. 
Balancing the lightweight nature and accuracy of simulator models is challenging, making it important to choose a model based on the algorithm's application.

For example, Sawada et al. base their work on the KT model for collision avoidance \cite{Sawada2021a}. In contrast, Maki et al. \cite{maki2020_1,maki2021_1}, Sawada et al. \cite{Sawada2021b,Sawada2023}, Miyauchi et al. \cite{Miyauchi2021PP}, Rachman et al. \cite{RACHMAN2022warm}, Wakita et al. \cite{Wakita2022RL}, and Suyama et al. \cite{Suyama2022} use the MMG model as the basis for their autonomous docking and undocking algorithms. Additionally, Bingham et al. \cite{Bingham2019} use Fossen's model for coupling evaluations of algorithms used in ocean robot contests and conducting competitions in a virtual environment.

In practical applications, there are examples of constructing dynamic models based on Fossen's model and developing control algorithms for the development of catamaran ASVs (Autonomous Surface Vehicles) \cite{Sarda2016}.

\paragraph{Models for embedding into model-based control laws or observers}
These dynamic models are used in control laws and observers. Nonlinear Kalman filters are employed to estimate unobservable states or model parameters directly.
For instance, Fossen et al. applied a dynamic model to an Extended Kalman Filter (EKF) and performed parameter identification based on trial results with actual ships \cite{fossen1996identification}.
From a control perspective, Fossen et al. used the dynamic model in an EKF to estimate yaw angle, yaw rate, and ship speed from only multiple GNSS position data. 
They conducted control simulations for heading keeping based on these estimated values \cite{fossen_2022}. Moreover, nonlinear control models, such as the backstepping method, have also been demonstrated \cite{Fossen1999tutorial}.

In previous works, the representation of fluid dynamic forces during low-speed navigation proposed by Karasuno was employed to control the laying of submarine cables using model predictive control \cite{Hamamatsu2008En}.
In this case, C/GMRES (continuation/GMRES) method \cite{OHTSUKA2004} was used for control. 
Recently, MMG has been directly applied in predictive ship-handling control during the approach phase of docking, achieving successful control of the large ferry and the cargo ship, as demonstrated by groups such as Ioki et al. \cite{Ioki2022_En} and Habu et al. \cite{Habu2022_En}. 
Contrary to this, simpler models, like Rachman's \cite{RACHMAN2022DPS}, have been used for docking scenarios.


\section{Categorization of the dynamic models and its histories}
\label{sec:history_categorization}



The history of research on dynamic models, categorized in this chapter, spans several purposes, reflecting a long history. Pre-war studies on ship maneuvering dynamics focused on topics such as the turning radius of warships. Particularly, in the case of the Imperial Japanese Navy, estimating the pivot point \cite{Akazaki1937En,Akazaki1968En} was the main focus of maneuvering studies. The analytical foundation based on motion equations and state equations, akin to the present, can be traced back to the pioneering work of Davidson and Schiff \cite{davidson1946}. Post-war in Japan, research in fluid dynamics by Motoyoshi \cite{Motora1959En} and others, stability analysis by Motoyoshi \cite{Motora1955En}, and control engineering research led by Inoue \cite{Inoue1979En} and Nomoto \cite{Nomoto1956_En} laid the groundwork for the comprehensive technological system of ship maneuvering as it is understood today.

In the analysis based on maneuvering motion equations, modeling fluid dynamic forces on the hull, propeller, and rudder has been a constant focus of research. 
Abkowitz \cite{Abkowitz1964} proposed a whole-ship model, expressing fluid dynamic forces as state variables. 
The MMG model introduced by Ogawa et al. \cite{Ogawa1977MMG_En,Yasukawa2015MMG} was groundbreaking, allowing for the representation of the ship components and their interactions.
When considering maneuvering motion, the primary focus is often on horizontal-plane motion. 
While studies on maneuvering motion in waves exist \cite{Nonaka1990En,Yasukawa2006wave1En,Yasukawa2006wave2En}, heave and pitch are often neglected.
In contrast, roll motion is frequently considered, especially in the case of high-speed vessels, with studies using a 4-degree-of-freedom (DoF) model \cite{Eda1980rolling,Son_Nomoto1981,Yasukawa2010RollEn}.
However, for the scope of this paper, which focuses on horizontal-plane motion, we describe the 3-DoF model.

With the development of motion equation modeling in the 1980s, dynamic models were classified into response models and fluid dynamics models \cite{Ogawa1977MMG_En,Nakato1981sympo_En}. Since then, various new dynamic models have been proposed. In this paper, considering the latest research trends, we categorize the currently practical models as follows.
\begin{description}
    \item[$\bullet$ Response model]\ \\
    A model derived by focusing onlf for the control inputs and the ship's response, without estimating each coefficient using hydrodynamics. However, in the case of models like Nomoto's KT model, it may also be derived directly from fluid dynamics models. This category includes:
    \begin{itemize}
        \item Nomoto's KT model
        \item AR models
        \item Neural network models
    \end{itemize}
    \item[$\bullet$ Fluid dynamics model] \ \\A model that estimates each coefficient using fluid dynamics while formulating and solving the maneuvering motion equations. This category includes:
        \begin{itemize}
            \item Whole-ship models
            \item Modular-type models
        \end{itemize}
\end{description}

Each of these models includes several coefficients, and an approach is taken to determine them through some means. We will now provide an overview of these models individually.

\subsection{Response models}
\subsubsection{Nomoto's KT model \cite{Nomoto1957En} and its extension}
In the case of the response model, Nomoto's pioneering research \cite{Nomoto1957En} is well-known. Taking the origin of the ship-fixed coordinate system at the center of gravity of the ship, the linear equations of motion for the sway angle $\beta$ and the rate of turn $r$ in the whole model format can be expressed as follows:
\begin{equation}
    \left\{
    \begin{aligned}
        &(m^{\prime}+m_{y}^{\prime})\left(\frac{L}{U}\right) \frac{d \beta}{d t}+C_{Y \beta} \cdot \beta\\
        &\quad-\left((m^{\prime}+m_{x}^{\prime})-C_{Y r}\right)\left(\frac{L}{U}\right) r=C_{Y \delta} \cdot \delta \\
        &(I_{zz}^{\prime}+J_{zz}^{\prime})\left(\frac{L}{U}\right)^{2} \frac{d r}{d t}+C_{N r}\left(\frac{L}{U}\right) r -C_{N \beta} \cdot \beta=C_{N \delta} \cdot \delta
    \end{aligned}
    \right.
\end{equation}
$C_{Yx}$ and $C_{Nx}$ represent the sway force and yaw moment acting on the ship, respectively, and their subscripts $x$ indicate the elements from which they originate. By eliminating one of the variables from this equation, a response model can be obtained. For example, considering the case of $r$, the following can be derived:
\begin{equation}
    \begin{gathered}
        T_{1} T_{2} \frac{\mathrm{d}^2 r}{\mathrm{d} t^2}+\left(T_{1}+T_{2}\right) \frac{\mathrm{d} r}{\mathrm{d} t}+r=K \delta+K T_{3} \frac{\mathrm{d} \delta}{\mathrm{d} t}\\
        \text{where} \, \left\{
        \begin{aligned}
            K &=\left(\frac{U}{L}\right) \frac{C_{N \beta} C_{Y \delta}+C_{Y \beta} C_{N \delta}}{C_{Y \beta} C_{N r}-\left(m_{x}-C_{Y r}\right) C_{N \beta}} \\
            \left(T_{1}+T_{2}\right) &=\left(\frac{L}{U}\right) \frac{m_{y} C_{N r}+n C_{Y \beta}}{C_{Y \beta} C_{N r}-\left(m_{x}-C_{Y r}\right) C_{N \beta}} \\
            T_{1} T_{2} &=\left(\frac{L}{U}\right)^{2} \frac{m_{y} \cdot n}{C_{Y \beta} C_{N r}-\left(m_{x}-C_{Y r}\right) C_{N \beta}} \\
            T_{3} &=\left(\frac{L}{U}\right) \frac{m_{y} C_{N \delta}}{C_{N \beta} C_{Y \delta}+C_{Y \beta} C_{N \delta}}
        \end{aligned}
        \right.
    \end{gathered}
\end{equation}
Furthermore, by assuming that the response related to the maneuvering motion is gradual, the one obtained by approximating it as a first-order system is called Nomoto's KT equation.
\begin{equation}
    T \frac{\mathrm{d}r}{\mathrm{d}t} + r = K \delta
\end{equation}
Z-test is a convenient and widely used method to determine the coefficients in the equation. Through this test, coefficients $T$ and $K$ are obtained, and these are given by:
\begin{equation}
    \left\{
    \begin{aligned}
        K^{\prime} &= K/(U/L)\\
        T^{\prime} &= T/(L/U)
    \end{aligned}
    \right.
\end{equation}

This allows for comparison with other vessels and facilitates direct discussion of the vessel's maneuvering performance. However, it is known that scale effects exist \cite{Okamoto1972En}. Various analytical methods have been proposed for addressing this issue \cite{Nomoto1969En}.

An extension of Nomoto's KT model considering non-linear terms was proposed by Norrbin in 1963 and takes the following form:
\begin{equation}
\label{eq:norrbin}
    T \frac{\mathrm{d}r}{\mathrm{d}t} + r + c_3 r^3= K \delta
\end{equation}

The study by Kim et al. \cite{Kim1978En} is known for determining the coefficients in the Eq (\ref{eq:norrbin}). It is interesting to note that the solution is not unique.

Efforts have been made to organize the coefficients of response models in terms of various parameters such as ship type, trim coefficients, and rudder area \cite{Nomoto1957En}. Referencing the study, it is possible to promptly create the required dynamic model for simulators based on past data. Additionally, model-based control using response models has been extensively conducted \cite{Fossen1999tutorial}.

\subsubsection{Auto Regressive (AR) model}
The Auto-Regressive (AR) model is a model that represents a time series $\{ x_{t}\}_{t=1}^{N} \in \mathbb{R}$ at a certain time point using past time series data as explanatory variables.
The model is essentially represented as Eq. \ref{eq_AR_model}:
\begin{equation}
    \label{eq_AR_model}
    x_{t}=\sum_{i=1}^{m} a_{i} x_{t-i}+v_{t}\ (t=m+1, \ldots, N)
\end{equation}
Here, $m \in \mathbb{N}$ represents the order of the AR model, a natural number. $a_{1}, \ldots, a_{m} \in \mathbb{R}$ are the auto-regressive coefficients of the AR model. These coefficients are determined by solving using the least squares method based on past time-series data. Additionally, the order of the regression model is often determined using the Akaike Information Criterion (AIC).
$v_t \in \mathbb{R}$ represents white noise with a mean of $0$ and a variance of $\sigma^2$. Åström and Källström use a probabilistic model in the form of an Auto-Regressive with eXogenous (ARX) model to express the motion due to steering and perform system identification \cite{astrom_kallstrom_1976}. Källström and others have also constructed an applied autopilot system using the ARX model \cite{kallstrom1979adaptive}. Otsu designed a control system AR model by introducing a term for control inputs, determining optimal control inputs from past time-series data, control inputs, and currently observed state variables. Otsu applied this method to an actual ship and performed heading control \cite{ohtsu_1980}. Additionally, discretizing Nomoto's first-order approximation model results in an ARX model. Iseki and others performed parameter identification of this ARX model using an Infinite Impulse Response (IIR) filter \cite{Iseki1998iir_En}. More recently, Jiang et al. used an AR model to investigate the effect of hull scale on real-time ship motion prediction \cite{JIANG2020107202}.

\subsubsection{Neural Network}
    An artificial neural network (ANN) is a model that can represent the relationship between control inputs and the ship's responses and is a multi-input, multi-output model with high fitting capabilities. ANN is a mathematical model inspired by the neural circuits of the brain and is constructed by combining multiple artificial neurons. Artificial neurons are processing units that pass a weighted sum of inputs through an activation function such as a sigmoid function or ReLU function, and are represented as follows:
    \begin{equation}
        \boldsymbol{y}=\boldsymbol{f}(\boldsymbol{W}\boldsymbol{x}+\boldsymbol{b})
        \label{eq:neuron}
    \end{equation}
    where $\boldsymbol{x} \in \mathbb{R}^{n}$ is the input vector, $\boldsymbol{y} \in \mathbb{R}^{m}$ is the output vector, $\boldsymbol{W} \in \mathbb{R}^{m \times n}$ is the weight matrix, $\boldsymbol{b} \in \mathbb{R}^{m}$ is the bias vector, and $\boldsymbol{g}: \mathbb{R}^{m} \rightarrow \mathbb{R}^{m}$ is the activation function. In an ANN, the adjustment of weights is done using training data to represent the relationship between input and output. A learning method using backpropagation is commonly used for this weight adjustment.
    
    For ship dynamic models, various methods have been proposed in which feed-forward neural networks (FNNs) and recurrent neural networks (RNNs) are used. 
    Moreira et al. \cite{MOREIRA2003,Moreira2012} proposed a maneuvering simulation model that recursively utilizes an FNN-based maneuvering model and validated it using data from maneuvering simulation of Mariner class ship and full-scale ship trials of a catamaran. Rajesh et al. \cite{RAJESH2008} identified the 3-DoF nonlinear steering model for the large tanker by expressing the nonlinear terms in terms of FNNs. Zhang et al. \cite{Zhang2013} identified a function representing nonlinear fluid dynamic forces using an FNN with Chebyshev orthogonal basis functions. Oskin et al. \cite{OSKIN2013} estimated the response model of yaw direction velocity to rudder angle input using RNN. Wakita et al. \cite{Wakita2022} estimated  3-DOF dynamic models using RNN and validated them using data from free-running model tests. RNNs can account for the effects of past sequences in time-series data by retaining internal states as memory, and several identification methods have been also proposed that use Long Short-Term Memory (LSTM), which can handle longer-term memory \cite{WOO2018,Jiang2022}.

\subsection{Model based on fluid dynamics}
The fluid dynamics model yields the following equations of motion in the ship-fixed coordinate system, with the origin at the midship of the hull.
\begin{equation}
    \begin{split}
        m \dot{u} - m v r -x_{\mathrm{G}}mr^{2} &= F_{X}\\
        m \dot{v} + m u r +x_{\mathrm{G}}m\dot{r} &= F_{Y}\\
        (I_{zz}+x_{\mathrm{G}}^{2}m) \dot{r} + x_{\mathrm{G}} m  ur &= M_{M} \enspace .
    \end{split}
    \label{eq:motion}
\end{equation}
In this paper, we classified the model based on how the fluid dynamic force terms on the right-hand side are solved. Specifically, we categorized the model into a modular-type model, where the fluid dynamic forces are decomposed for each component of the ship, namely the hull, propeller, and rudder, and a whole-ship model, where the fluid dynamic forces are not separated and the entire ship is treated as one system.

\subsubsection{Whole-ship model}\label{sec:wbmodel}
The whole-ship model treats the right-hand side of \cref{eq:motion} as a single system without decomposing the contributions of individual modules of the ship (hull, propeller, rudder, etc.).

The most widely used white-box model of whole-ship models is the Abkowitz model \cite{Abkowitz1964}. This model assumes perturbation motion caused by small rudder angle change $\Delta\delta$ from a state $\mathbf{x}_{0}$ in which the ship with constant forward speed $U$ and constant propeller rotation. Then, fluid dynamic forces \cref{eq:abko_variables} are assumed as the function of the state variables, control inputs, and their time derivatives:
\begin{equation}\label{eq:abko_variables}
    \begin{split}
        F_{X} &= f(\mathbf{x}) \\
        F_{Y} &= g(\mathbf{x}) \\
        M_{M} &= h(\mathbf{x}) \enspace ,
        \end{split}
\end{equation}
where
\begin{align}
       \mathbf{x}         &= (u,~v,r,~\dot{u},~\dot{v},\dot{r},\delta,~\dot{\delta})^{\top} \\
        \mathbf{x}_{0}    &= ( U,~0,~0,~0,~0,~0,~0,~0)^{\top} \enspace,
\end{align}
and the right-hand side of \cref{eq:motion} is expanded as Taylor series.
Based on the transverse symmetricity of the ship hull, a polynomial is determined by selecting its terms. Moreover, as the user has the flexibility to choose the maximum degree of the polynomial, there are several derivative types available (for example, see \cite{Chislett1965}). Here, as an example, the polynomial from Abkowitz's work \cite{Abkowitz1980} is presented by expanding \Cref{eq:abko_variables} up to the third order in $\mathbf{x}$.
\begin{equation}
    \begin{aligned}
    &F_{X}(\mathbf{x}) \approx X\left(\mathbf{x}_{0}\right)\\&+\sum_{i=1}^{n}\left(\left.\frac{\partial X(\mathbf{x})}{\partial x_{i}}\right|_{\mathbf{x}_{0}} \Delta x_{i}+\left.\frac{1}{2} \frac{\partial^{2} X(\mathbf{x})}{\left(\partial x_{i}\right)^{2}}\right|_{\mathbf{x}_{0}} \Delta x_{i}^{2}+\left.\frac{1}{6} \frac{\partial^{3} X(\mathbf{x})}{\left(\partial x_{i}\right)^{3}}\right|_{\mathbf{x}_{0}} \Delta x_{i}^{3}\right) \\
    &F_{Y}(\mathbf{x}) \approx Y\left(\mathbf{x}_{0}\right)\\&+\sum_{i=1}^{n}\left(\left.\frac{\partial Y(\mathbf{x})}{\partial x_{i}}\right|_{\mathbf{x}_{0}} \Delta x_{i}+\left.\frac{1}{2} \frac{\partial^{2} Y(\mathbf{x})}{\left(\partial x_{i}\right)^{2}}\right|_{\mathbf{x}_{0}} \Delta x_{i}^{2}+\left.\frac{1}{6} \frac{\partial^{3} Y(\mathbf{x})}{\left(\partial x_{i}\right)^{3}}\right|_{\mathbf{x}_{0}} \Delta x_{i}^{3}\right) \\
    &M_{M}(\mathbf{x}) \approx N\left(\mathbf{x}_{0}\right)\\&+\sum_{i=1}^{n}\left(\left.\frac{\partial N(\mathbf{x})}{\partial x_{i}}\right|_{\mathbf{x}_{0}} \Delta x_{i}+\left.\frac{1}{2} \frac{\partial^{2} N(\mathbf{x})}{\left(\partial x_{i}\right)^{2}}\right|_{\mathbf{x}_{0}} \Delta x_{i}^{2}+\left.\frac{1}{6} \frac{\partial^{3} N(\mathbf{x})}{\left(\partial x_{i}\right)^{3}}\right|_{\mathbf{x}_{0}} \Delta x_{i}^{3}\right) \enspace.
\end{aligned}
\end{equation}
By organizing the obtained polynomial based on certain assumptions and rearranging the terms, the Abkowitz model in \cref{eq:abkowitz} can be derived, with the acceleration-related terms on the left side and the rest on the right side.
\begin{align}\label{eq:abkowitz}
    \begin{aligned}
        &\dot{u}=\frac{f_{1}(u, v, r, \delta)}{\left(m-X_{\dot{u}}\right)} \\
        &\dot{v}=\frac{\left(I_{z}-N_{\dot{r}}\right) f_{2}(u, v, r, \delta)-\left(m x_{\mathrm{G}}-Y_{\dot{r}}\right) f_{3}(u, v, r, \delta)}{\left(m-Y_{\dot{v}}\right)\left(I_{z}-N_{\dot{r}}\right)-\left(m x_{\mathrm{G}}-N_{\dot{v}}\right)\left(m x_{\mathrm{G}}-Y_{\dot{r}}\right)} \\
        &\dot{\mathrm{r}}=\frac{\left(m-Y_{\dot{v}}\right) f_{3}(u, v, r, \delta)-\left(m x_{\mathrm{G}}-N_{\dot{v}}\right) f_{2}(u, v, r, \delta)}{\left(m-Y_{\dot{v}}\right)\left(I_{z}-N_{\dot{r}}\right)-\left(m x_{\mathrm{G}}-N_{\dot{v}}\right)\left(m x_{\mathrm{G}}-Y_{\dot{r}}\right)} \enspace,
        \end{aligned}
\end{align}
where
\begin{align}\label{eq:abkowitz_force}
    \begin{aligned}
f_{1}(u, v, r, \delta)&=X_{*}+X_{u} \Delta u+X_{u u} \Delta u^{2}+X_{u u u} \Delta u^{3}+X_{v v} v^{2}\\
            &+\left(X_{r r}+m X_{\mathrm{G}}\right) r^{2}+X_{\delta \delta} \delta^{2}+X_{v v u} v^{2} \Delta u\\
            &+X_{r r u} r^{2} \Delta u+X_{\delta \delta u} \delta^{2} \Delta u+\left(X_{v r}+m\right) v r+X_{v \delta} v \delta\\
            &+X_{r \delta}^{r \delta}+X_{v r u} v r \Delta u+X_{v \delta u} v \delta \Delta u+X_{r}{ }_{u} r \delta \Delta u\\
        f_{2}(u, v, r, \delta)&=Y_{*}+Y_{*u} \Delta u+Y_{*u} \Delta u^{2}+Y_{v} v+Y_{v v v} v^{3}+Y_{v r r} {v r^{2}}\\
            &+Y_{v \delta \delta}v \delta^{2}+Y_{v u} v \Delta u+Y_{v u u} v \Delta u^{2}+\left(Y_{r}-m u\right) r\\
            &+Y_{r r r} r^{3}+Y_{r v v} rv^{2}+Y_{r \delta \delta}r \delta^{2}+Y_{r u} \Delta u+Y_{r u u} r \Delta u^{2}\\
            &+Y_{\delta} \delta+Y_{\delta \delta \delta} \delta^{3}+Y_{\delta v v} \delta v^{2}+Y_{\delta r r} \delta r^{2}+Y_{\delta u} \delta \Delta u\\
            &+Y_{\delta u u} \delta \Delta u^{2}+Y_{v r \delta} v r \delta\\
        f_{3}(u, v, r, \delta)&=N_{*}+N_{*u} \Delta u+N_{*uu} \Delta u^{2}+N_{v} v+N_{v v v} v^{3}\\
            &+N_{v r r} v r^{2}+N_{v \delta \delta}{v \delta^{2}}+N_{v u} v \Delta u+N_{v u u} v \Delta u^{2}\\
            &+\left(N_{r}-m_{\mathrm{G}} u\right) r+N_{r r r} r^{3}+N_{r v v}r v^{2}+N_{r \delta \delta} r\delta^{2}\\
            &+N_{r u} r \Delta u+N_{r u u} r \Delta u^{2}+N_{\delta} \delta+N_{\delta \delta \delta} \delta^{3}+N_{\delta v v} \delta v^{2}\\
            &+N_{\delta r r} \delta r^{2}+N_{\delta u} \delta \Delta u+N_{\delta u u} \delta \Delta u^{2}+N_{v r \delta}v r \delta \enspace.
\end{aligned}
\end{align}

The Abkowitz model is commonly used as a nonlinear model that includes higher-order terms, but in Abkowitz's literature \cite{Abkowitz1964}, linearized motion equations and stability analysis based on them are also presented. The drawbacks of the Abkowitz model include the unclear physical interpretation of higher-order terms, limitations arising from assuming small variations from a constant speed $U$, making it challenging to model significant speed reductions in maneuvers with large rudder angles, and the inability to incorporate variations in propeller revolutions and the influence of wind forces into motion simulations.

\subsubsection{Modular-type model}
As modular-type models, we introduce the following three:
\begin{enumerate}
    \item MMG model
    \item Modified Abkowitz model
    \item Fossen's model
\end{enumerate}

\paragraph{MMG model}
The MMG (Maneuvering Modeling Group) model was developed in 1976 to address the following weaknesses of polynomial-type whole-ship models.
\begin{itemize}
    \item In the representation of fluid dynamic forces by polynomials, the values of each coefficient vary depending on the choice of polynomial terms. This poses a challenge for comparing results between research institutions.
    \item It cannot accommodate partial design changes, such as alterations to the rudder.
    \item It cannot correlate with actual ship behavior.
    \item The physical meaning of the coefficients of the polynomial is unclear.
\end{itemize}
In 1976, the fundamental concepts of the MMG (Maneuvering Modeling Group) model were developed \cite{Ogawa1977MMG_En} and further discussed in \cite{Ogawa1981sympo_En}.
Afterward, upon closer examination of the details of the dynamic models widely used in Japan, slight variations were observed among different institutions. To standardize these models, the research committee on standardization of ship maneuverability prediction models \cite{P29report_En} was established. The summary paper \cite{Yasukawa2015MMG} serves as an overview, providing a detailed mathematical representation of the standardized MMG model. However, it is important to note that the standardized MMG model assumes that the lateral flow velocity is significantly smaller than the forward speed, and it does not encompass maneuvers such as berthing and unberthing operations or propeller reversal maneuvers.


In the MMG model, $F_X$, $F_Y$, and $M_M$ in \Cref{eq:motion} is divided into the velocity-dependent components $X_S$, $Y_S$, and $N_S$ and the acceleration-dependent hydrodynamic component $X_A$, $Y_A$, and $N_A$. The steady fluid dynamic forces dependent on velocity components are expressed by considering the individual actions of the hull, propeller, rudder, etc., and the fluid dynamic forces resulting from their mutual interference effects.

By employing this representation, it becomes possible to theoretically or experimentally investigate the individual characteristics of the hull, propeller, and rudder, as well as the interference effects among them. Furthermore, due to the clear physical meaning of each term in the model, it is relatively straightforward to incorporate scale effects between full-scale and model-scale ships, account for shallow water effects, and introduce external disturbances.

In the MMG model, $X_S$, $Y_S$, and $N_S$ are generally expressed as follows:
\begin{equation}\label{eq:mmg_decomposition}
    \begin{split}
        X_S &= X_{\mathrm{H}} + X_{\mathrm{P}} + X_{\mathrm{R}} + X_{\mathrm{T}} + X_{\mathrm{wind}} + X_{\mathrm{wave}} \\
        Y_S &= Y_{\mathrm{H}} + Y_{\mathrm{P}} + Y_{\mathrm{R}} + Y_{\mathrm{T}} + Y_{\mathrm{wind}} + Y_{\mathrm{wave}} \\
        N_S &= N_{\mathrm{H}} + N_{\mathrm{P}} + N_{\mathrm{R}} + N_{\mathrm{T}} + N_{\mathrm{wind}} + N_{\mathrm{wave}} \enspace .
\end{split}
\end{equation}

In the subscripts within the equations, $\mathrm{H}$ represents the fluid dynamic forces related to the hull, $\mathrm{R}$ corresponds to the fluid dynamic forces related to the rudder, $\mathrm{P}$ denotes the fluid dynamic forces related to the propeller, $\mathrm{T}$ signifies the fluid dynamic forces associated with the side thruster, $\mathrm{wind}$ stands for the fluid dynamic forces related to the wind, and $\mathrm{wave}$ represents the fluid dynamic forces related to waves.

Many other models can be derived from this model. As previously mentioned, Nomoto's KT model is obtained by linearizing this model and approximating it as a first-order system.

The acceleration-dependent fluid dynamic forces $X_A$, $Y_A$, and $N_A$ follow Lamb's \cite{Lamb1932} approach and are expressed as follows. However, differences arise in dynamic models, such as when eliminating certain terms for a ship with fore-and-aft symmetry or incorporating terms that are difficult to separate in velocity-dependent steady fluid dynamic forces in tank tests into the velocity-dependent steady fluid dynamic forces.
\begin{equation}
    \begin{split}
        X_A &=  -m_x \dot{u} + m_y v r +  m_y \alpha_y r^2 \\
        Y_A &= -m_y \dot{v} - m_x u r -  m_y \alpha_y \dot{r}  \\
        N_A &=  -J_{zz} \dot{r} - m_y \alpha_y (\dot{v} + u r) - (m_y -m_x) u v \enspace .
    \end{split}
\end{equation}


Here, the added masses of surge and sway motion and the added moment of inertia of yaw motion are denoted by $m_x$, $m_y$ and $I_z$, respectively.
While there is currently limited detailed literature on modeling for low-speed navigation, one notable example is the paper by Miyauchi\cite{Miyauchi2022SI}. The model representing low-speed maneuvering typically exhibits the following characteristics compared to the conventional MMG model.

Firstly, let's discuss the fluid dynamic forces acting on the hull. The non-dimensional yaw rate $r'$ is often expressed as $r' = rL/U$ in the usual range of ship speeds. When the cubic term of $r'$ is dimensionalized, it becomes as follows:
\begin{equation}
        \frac{1}{2}\rho LdU^2 \{ Y_{rrr}^{\prime} r^{\prime 3} \}
        =\frac{1}{2}\rho L^2dY_{rrr}^{\prime}r^2 \frac{Lr}{U}
\end{equation}
In low-speed maneuvering, situations resembling in-place turning with $U=0$ can occur, making it challenging to compute under such circumstances. Therefore, models based on the concept of cross-flow drag \cite{Yoshimura2009a} or dimensionless parameters like $r' = r/\sqrt{g/L}$ \cite{kose1984_En} have been proposed to address these challenges.

Typically, in the usual range of ship speeds, the forward speed dominates over the lateral flow velocity, and the sideslip angle $\beta$ is assumed to be within approximately 20 degrees. Hence, assumptions such as $u \backsimeq U$ are often employed. However, it is crucial to account for cases where large sideslip angles occur, especially during lateral movements.

Furthermore, models for propellers and rudders need to consider various scenarios, such as high or low propeller loading in self-propulsion conditions and the possibility of propeller reverse rotation. Accurately representing the forces generated by the propeller and rudder under such conditions is essential.

\paragraph{Modified Abkowitz model}
Besides the MMG model, another Modular-type model is proposed by Abkowitz and colleagues \cite{Hwang1980,Abkowitz1980}. This model, referred to as the modified Abkowitz model in this paper, is an advancement of the Abkowitz model presented in Section 4.2 \cref{sec:wbmodel}. It was introduced to enhance the estimation performance of the Abkowitz model through system identification.

The modified Abkowitz model replaces \cref{eq:abkowitz_force} with the following equation:
\begin{align}
   & \begin{aligned}
    f_{1} &=\eta_{1}^{\prime}\left[\frac{\rho}{2} L^{2}\right] u_{r}^{2}+\eta_{2}^{\prime}\left[\frac{\rho}{2} L^{3}\right] n u_{r}+\eta_{3}^{\prime}\left[\frac{\rho}{2} L^{4}\right] n^{2} \\
    &-C_{R}\left[\frac{\rho}{2} S u_{r}^{2}\right]+X_{v_{r}^{2}}^{\prime}\left[\frac{\rho}{2} L^{2}\right] v_{r}^{2}+X_{e^{2}}^{\prime}\left[\frac{\rho}{2} L^{2} c^{2}\right] e^{2} \\
    &+\left(X_{r}^{\prime}+m^{\prime} x_{G}^{\prime}\right)\left[\frac{\rho}{2} L^{4}\right] r^{2}+\left(X_{v_{r}}^{\prime}+m^{\prime}\right)\left[\frac{\rho}{2} L^{3}\right] v_{r} r \\ &+X_{v^{2}_{r}r^{2}}^{\prime}\left[\frac{\rho}{2} L^{4}U_{r}^{-2}\right] v^{2}_{r} r^{2} 
    \end{aligned}\\
    &\begin{aligned}
        f_{2} &=Y_{0}^{\prime}\left[\frac{\rho}{2} L^{2}\left(\frac{u_{A \infty}}{2}\right)^{2}\right]\\
        &+\left\{Y_{v_{r}}^{\prime}\left[\frac{\rho}{2} L^{2} U_{r}\right] v_{r}
        +Y_{\delta}^{\prime}\left(c-c_{0}\right) \frac{\rho}{2} L^{2} v_{r}\right\} \\
        &+\left\{\left(Y_{r}^{\prime}-m^{\prime} u_{r}^{\prime}\right)\left[\frac{\rho}{2} L^{3} U_{r}\right] r -\frac{Y_{\delta}^{\prime}}{2}\left(c-c_{0}\right) \frac{\rho}{2} L^{3} r\right\} \\
        &+Y_{\delta}^{\prime}\left[\frac{\rho}{2} L^{2} c^{2}\right] \delta \\
        &+Y_{v_{r}^{3}}^{\prime}\left[\frac{\rho}{2} L^{2} U_{r}^{-1}\right] v_{r}^{3}+Y_{v_{r}}^{\prime}\left[\frac{\rho}{2} L^{3} U_{r}^{-1}\right] v_{r}^{2} r \\
        &+Y_{r^{2} v_{r}}^{\prime}\left[\frac{\rho}{2} L^{4} U_{r}^{-1}\right] r^{2} v_{r}+Y_{r}^{\prime}\left[\frac{\rho}{2} L^{5} U_{r}^{-1}\right] r^{3} \\
        &+Y_{e^{3}}^{\prime}\left[\frac{\rho}{2} L^{2} c^{2}\right] e^{3}
    \end{aligned} \\
    &\begin{aligned}
        f_{3}&=N_{0}^{\prime}\left[\frac{\rho}{2} L^{3}\left(\frac{u_{A \infty}}{2}\right)^{2}\right]\\
        &+\left\{N_{v_{r}}^{\prime}\left[\frac{\rho}{2} L^{3} U_{r}\right] v_{r}
        -N_{\delta}^{\prime}\left(c-c_{0}\right) \frac{\rho}{2} L^{3} v_{r}\right\}\\
        &+\left\{\left(N_{r}^{\prime}-m^{\prime} x_{G}^{\prime} u_{r}^{\prime}\right)\left[\frac{\rho}{2} L^{4} U_{r}\right] r
        +\frac{1}{2} N_{\delta}^{\prime}\left(c-c_{0}\right) \frac{\rho}{2} L^{4} r\right\}\\
        &+N_{\delta}^{\prime}\left[\frac{\rho}{2} L^{3} c^{2}\right] \delta\\
        &+N_{v_{r}{ }^{3}}\left[\frac{\rho}{2} L^{3} U_{r}^{-1}\right] v_{r}^{3}+N_{v_{r}}^{{ }_{r}}{ }_{r}\left[\frac{\rho}{2} L^{4} U_{r}^{-1}\right] v_{r}^{2} r\\
        &+N_{r}^{\prime} 2_{v_{r}}\left[\frac{\rho}{2} L^{5} U_{r}^{-1}\right] r^{2} v_{r}+N_{r}^{\prime}\left[\frac{\rho}{2} L^{6} U_{r}^{-1}\right] r^{3}\\
        &+N_{e^{3}}^{\prime}\left[\frac{\rho}{2} L^{3} c^{2}\right] e^{3}
        \end{aligned}
\end{align}
This model incorporates thrust and hull resistance separation using thrust coefficients $\eta_{1}^{\prime},~\eta_{2}^{\prime},~\eta_{3}^{\prime}$. Furthermore, it expresses the rudder force, previously represented in the Abkowitz model as a coupling term between $\delta$ and $v,~r$, in terms of the effective inflow angle $e$ and the weighted average inflow velocity $c$. The subscript $r$ indicates the relative velocity considering tidal flow. This allows for the accommodation of large rudder angles and propeller reversal. However, unlike the MMG model, it is not considered a Modular-type model in the sense that it does not assume assembly from the individual performance of the rudder and propeller. It can still be valuable for partial design considerations.

\paragraph{Fossen's Matrix-vector Representation model}
Fossen's model, as presented in \cite{fossen2021handbook}, is characterized by expressing the equations of motion in a matrix-vector form. Adopting a linearized system in matrix form is convenient for stability analysis.

In Fossen's model, the state variables are conventionally represented as $\boldsymbol{\nu}=(u,~v,~r)^{\top} \in \mathbb{R}^3$, and the equations of motion are expressed in vector form with respect to $\boldsymbol{\nu}$.
\begin{equation}\label{eq:fossen's_model}
    \boldsymbol{M\dot{\nu}} + \boldsymbol{C(\nu)\nu} + \boldsymbol{D(\nu)\nu}  = \boldsymbol{\tau + \tau}_{\text{wind}} + \boldsymbol{\tau}_{\text{wave}}
\end{equation}

Here, $\boldsymbol{M}$ is the inertia matrix, including added mass, $\boldsymbol{C}$ is the Coriolis and centripetal force matrix, $\boldsymbol{D}$ is the damping matrix, $\boldsymbol{\tau}$ is the control force vector, and $\boldsymbol{\tau}_{\mathrm{wind}},~\boldsymbol{\tau}_{\mathrm{wave}}$ are external force vectors due to wind and waves. Users have the flexibility to choose the matrix representations, and Fossen's textbook \cite{fossen2021handbook} provides examples of various damping matrices, including those using linear matrices, cross-flow drag, a second-order polynomial with absolute value function, and a third-order polynomial based on \cite{Abkowitz1964}.

For control design, this model is often linearized with respect to control inputs $\mathbf{u}$ as $\boldsymbol{\tau}=\mathbf{B}\mathbf{u}$. For instance, for azimuth thrusters, the modeling is illustrated as follows \cite{fossen_2022}. The control force $\boldsymbol{\tau}$ in \cref{eq:fossen's_model} is defined as follows:
\begin{gather}
    \begin{aligned}
        &\boldsymbol{\tau} = (1-t)T_{|n|n}|n|n
        \begin{bmatrix}
            \cos(\alpha) \\
            \sin(\alpha) \\
            l_{x}\sin(\alpha)-l_{y}\cos(\alpha)
        \end{bmatrix}
        \\
        &-\boldsymbol{d}_{\mathrm{loss}}(n,~\alpha)u_{r} \enspace,
    \end{aligned}
\end{gather}
Here, $\boldsymbol{d}_{\mathrm{loss}}$ represents the loss term due to velocity-dependent propeller efficiency, and $\alpha$ is the azimuth angle. The mounting position of the azimuth thruster is denoted by $(l_{x},~l_{y})$. The input $\mathbf{u}$ is defined as follows:
\begin{gather}
    \mathbf{u} = 
    \begin{bmatrix}
        u_{1} \\
        u_{2}
    \end{bmatrix} =
    \begin{bmatrix}
         (1-t)T_{|n|n}|n|n\cos(\alpha) \\
         (1-t)T_{|n|n}|n|n\sin(\alpha)
    \end{bmatrix}\enspace,
\end{gather}
Terms in $\boldsymbol{\tau}$ can be summarized as:
\begin{equation}
    \tau = \boldsymbol{B}\mathbf{u} -\boldsymbol{D}_{\mathrm{loss}}(n,~\alpha)\boldsymbol{\nu}_{r}\enspace,
\end{equation}
where $\boldsymbol{B}$ and$\boldsymbol{D}_{\mathrm{loss}}$ can be defined as:
\begin{gather}
    \mathbf{B} = \begin{bmatrix}
        1 & 0\\
        0 & 1 \\
        -l_{t} & l_{x}
    \end{bmatrix},~ 
    \boldsymbol{D}_{\mathrm{loss}}(n,~\alpha) = (\boldsymbol{d}_{\mathrm{loss}}, ~ \boldsymbol{0}_{3\times2}) \enspace,
\end{gather}
Transferring the loss term to the left side yields the following:
\begin{equation}
    \begin{aligned}
        &\boldsymbol{M\dot{\nu}_{r}} + \boldsymbol{C(\nu_{r})\nu_{r}} + \big\{
        \boldsymbol{D}(\boldsymbol{\nu}_{r})+\boldsymbol{D}_{\mathrm{loss}}(n,~\alpha)\big\}\boldsymbol{\nu}_{r}\\
        &= \boldsymbol{B}\mathbf{u} + \boldsymbol{\tau}_{\text{wind}} + \boldsymbol{\tau}_{\text{wave}} \enspace.
    \end{aligned}
\end{equation}
Here, the above equation is for the relative velocity to seawater $\boldsymbol{\nu_{r}}$.

For the vessel having the usual rudder and propeller, there exists the following modeling:
Fossen \cite{fossen_2022} assume the force and moment $\tau$ generated by the actuator in \Cref{eq:fossen's_model}
\begin{gather}
    \tau  = 
    \begin{bmatrix}
        \tau_{1} \\
        \tau_{2} \\
        \tau_{6}
    \end{bmatrix} =
    \begin{bmatrix}
        -X_{\delta\delta}\delta^{2}+(1-t)T \\
        -Y_{\delta}\delta\\
        -N_{\delta}\delta
    \end{bmatrix}\enspace ,
\end{gather}
By using the following relation:
\begin{equation*}
    \left\{\begin{aligned}
        U_{\mathrm{R}} &\approx u_{\mathrm{R}}\\
        \alpha_{\mathrm{R}} &\approx\delta
    \end{aligned} \right.
\end{equation*}
then, the force and moment generated $\boldsymbol{\tau}_{\mathrm{R}}$ by rudder are approximated as following:
\begin{equation}
    \begin{aligned}
        \boldsymbol{\tau}_{\mathrm{R}}
        &=\left[\begin{array}{c}
        -\frac{1}{2}\left(1-t_{\mathrm{R}}\right) \rho U_{\mathrm{R}}^2 A_{\mathrm{R}} C_N \sin ^2(\delta) \\
        -\frac{1}{4}\left(1+a_{\mathrm{H}}\right) \rho U_{\mathrm{R}}^2 A_{\mathrm{R}} C_N \sin (2 \delta) \\
        -\frac{1}{4}\left(x_{\mathrm{R}}+a_{\mathrm{H}} x_{\mathrm{H}}\right) \rho U_{\mathrm{R}}^2 A_{\mathrm{R}} C_N \sin (2 \delta)
        \end{array}\right] \\
        &\approx\left[\begin{array}{c}
        -X_{\delta \delta} \delta^2 \\
        -Y_\delta \delta \\
        -N_\delta \delta
        \end{array}\right]\enspace.
    \end{aligned}
\end{equation}

Here, $(1-t_{\mathrm{R}})$, $a_{\mathrm{H}}$, and $x_{\mathrm{H}}$ are estimated by Kijima's regression formula \cite{kijima1990manoeuvring}. Further, the longitudinal effective velocity $u_{\mathrm{R}}$ at the rudder is estimated by the following expression used in the MMG model:
\begin{equation}
    \begin{aligned}
        u_{\mathrm{R}}=&\varepsilon u\left(1-w_P\right)\\
        &\cdot \sqrt{\eta\left(1+\kappa\left(\sqrt{1+\frac{8 K_T}{\pi J_a^2}}-1\right)\right)^2+(1-\eta)}
    \end{aligned}
\end{equation}
Fujii's formula \cite{Fujii_rudder_coef} is used for the rudder lift gradient coefficient $C_{N}$.

\subsection{Examples for other vehicles}
The discussion thus far has primarily focused on the three-degree-of-freedom dynamic model of Surge, Yaw, and Sway. However, in order to represent the turning motion of underwater vessels and high-speed planing boats, a six-degree-of-freedom dynamic model becomes necessary. For instance, Ghadimi \cite{Ghadimi2013} formulated the equations of motion with six degrees of freedom to analyze the behavior of planing boats in waves. The model decomposes loads into hydrostatic, hydrodynamic, wave force, and moment components to construct a comprehensive representation. Additionally, Kerdraon \cite{Kerdraon2020}, in the context of motion analysis for racing yachts in open seas, based the model on Fossen's work. The right-hand side loads in the model account for the forces from the hull, appendages, aerodynamics, and gravity.  Equations of motion with two-degree-of-freedom for heave and pitch motion, and with three-degree-of-freedom for surge, heave and pitch motion, are often used for porpoising, which is one of the main topics in planing hull stability \cite{Savitsky1964,Troesch1992,Hicks1995,Hamada2023}. Such an approach, where an appropriate degree of freedom is selected for the specific target, followed by the consideration of loads, is widely employed, as exemplified by these studies.

Regarding underwater vessels, there has been a strong emphasis on vertical stability analysis, particularly due to the constraint of not exceeding the depth barrier. Consequently, analyses using the Whole model appear to be prevalent. In such investigations, six-degree-of-freedom equations of motion, as exemplified in literature \cite{Gertler1967,Feldman1979}, are commonly employed. Vertical Planer Motion Mechanism (VPMM) tests, which involve vertical motion, have a long history, as seen in works such as \cite{Gertler1967PMM}. Some of the authors have also introduced devices used in these tests \cite{maki2018UUV}. Experimental formulas for various coefficients have been published since the 1950s, as evidenced by studies like \cite{Landweber1951,whicker1958,dempsey1977}. For instance, in Japan, a series of studies by Murakami \cite{Murakami1975En,Murakami1978En,Murakami2008En} are well-known. The works of Tokugawa and Kito \cite{Tokugawa1954En} are also widely recognized in this context.
%

\section{Trade-off between complexity and cost of building dynamic models}\label{sec:modeling_cost}

The previous chapter outlined the different types of dynamic models. For these dynamic models, tank tests, numerical calculations, and real and model ship tests are required to construct them. These require time and monetary costs.
While such costs are generally unwelcome, users also generally expect high model accuracy. There is generally a trade-off between the two. For instance, when constructing dynamic models based on fluid dynamics using captive model tests, the following costs are associated:

\begin{itemize}
    \item Since the model has a hydrodynamic theoretical background, knowledge of the model is necessary for model selection and experimental acquisition of model parameters. Therefore, before developing an algorithm, a person with knowledge of the model is required to build the model for each specific vessel.
    \item In order to conduct captive model tests, dedicated test facilities must be used. Therefore, control system manufacturers and classification societies that do not own test facilities face certain hurdles in using them. In addition, there are probably even fewer tanks available for shallow-water testing.
    \item As discussed below, the number of tests generally increases, especially when models are constructed for complex maneuvers at low speeds or in shallow water. Therefore, even institutions with dedicated testing facilities may find it time-consuming to create models for individual vessels.
\end{itemize}

The dynamic model required for the research and development of control systems is essentially different from the dynamic model required for design studies in the hull design phase. In the hull design phase, the dynamic model should be of a modular type because the dynamic model reflects design changes in the hull and appendage. However, in the research and development of control systems, the model doesn't have to be of a modular-type, since the hull, engine, and electrical hardware are often given. Rather, it is important that the dynamic model be a linear system or an input-affine system, and that the cost of generating the dynamic model for each ship be low.


Based on the authors' research and other studies, this chapter summarizes the accuracy required for dynamic models and the cost of constructing dynamic models for each method.

Firstly, let's organize the required accuracy for the models. For instance, the precision required for collision avoidance maneuvers may be relatively low. On the other hand, high precision may be necessary for estimating the behavior during emergency stops or berthing/unberthing maneuvers. When considering the latter, particularly during berthing/unberthing, the properties for which accuracy is crucial can be summarized as the following:

\begin{enumerate}
    \item Position, velocity, and yaw angle during sway motion, specifically crabbing, in shallow waters or near a quay
    \item Change of ship's velocities when the propeller revolution is reversed.
    \item Position (including kick), velocity, yaw angle (and its rate) during turning motion when a large rudder angle is taken. 
    \item Response characteristics and transient behavior of actuators, including factors such as dead time and delay.
    \item Noise and the discretization of measurements in the data acquisition process or control stage.
\end{enumerate}

The required estimation accuracy of the model depends on the objective and the object to be evaluated. The required accuracy criterion could be determined from the magnitude of the error in the motion between the simulation and the actual vessel. For instance, if we are considering speed reduction control before pier arrival, the accuracy of estimating the braking distance at pier arrival would be the target of evaluation. 
In this case, the distance between the final berthing position and the quay would determine the upper limit of the accuracy of the reproduction of the braking distance. If the distance from the defined stopping position to the quay is, for instance, 50 m, then the accuracy of the estimation of the braking distance must be sufficiently smaller than this distance.

Indeed, in the research on automatic berthing control by Sawada et al., it is mentioned that a position accuracy of approximately 1 meter or less is required for the sensor's target position \cite{Sawada2021b}. Similar discussions can be extended to simulations, and then the required accuracy (error between simulation and actual ship behavior) can be defined.

When discussing dynamic models, it is necessary to consider not only the form of the model but also how to obtain the parameters it contains. Various typical methods are listed in the ITTC\cite{ITTC2021manuvalidation}. In this section, we consider the cost of these methods.

\begin{enumerate}
    \item Use the parameter sets of existing type ships.
    \item Determine the parameter set from a regression formula that is generated from the databases for various ships.
    \item Determine the parameter set from the captive model tests or captive model tests using Computational Fluid Dynamics (CFD) for individual vessels.
    \item Determine the parameter set from the System Identification (SI) for free-running model test or actual ship operational data or CFD direct maneuvering simulations.
\end{enumerate}
The authors would also like to add a note about CFD, which also appeared at the end of the list. With recent improvements in computational power, the method of estimating motion by direct maneuvering simulation using CFD has also been used \cite{Carrica2013a,Chase2013}. In this method, the ship is considered a rigid body, and the instantaneous flow field around the ship is analyzed by the unsteady Raynolds-Averaged Navier-Stokes method to estimate the forces acting on the ship. The unsteady motion of the ship is then estimated by performing a coupled rigid-body and fluid calculation. Although this method can estimate the motion with high accuracy, the computational cost is very high. Therefore, it is not considered suitable for simulator applications at this time. Therefore, we decided not to discuss it in this paper.

\subsection{Direct application of type ship values}
With the use of the same or slightly modified coefficients as for a similar already existing ship (type ship), a simulator can be constructed. It should be noted, however, that although this method appears to be costless, some cost has already been paid in the modeling of the type ship.

\subsection{Determine hydrodynamic derivatives from database or regression model based on the database}
In this method, the coefficients are determined by using regression equations based on the ship's main requirements and so on. Such methods include, for example, just listing the MMG model: Motora's chart \cite{Motora1959En,Motora1960En,Motora1960_2En}, Inoue's equation \cite{Inoue1979En}, Kijima's equation \cite{kijima1990manoeuvring}, Yoshimura's formula\cite{Yoshimura2009a,Yoshimura2011_En}, Hasegawa's formula for $X_{vr}$ \cite{Hasegawa1980}, Fujiwara's formula for wind pressure \cite{Fujiwara1998_En}. There are several other examples. Of course, there are many other methods.


In general, many maneuvering motion model parameters need to be estimated for low-speed maneuvering motions during berthing/unberthing maneuvers. Unfortunately, no chart is currently available that can estimate those parameters

While MMG models can estimate motion with some accuracy, criticisms also have existed. For example, in the literature \cite{HUANG2020451}, it is stated that the coding cost of the MMG model is high. However, the adoption of MMG does not necessarily mean the high coding cost of the model. This is because a large amount of past testing data has already been accumulated for MMG models. Therefore, users can obtain models with a certain degree of accuracy at a low cost by referring to those existing results. This is due to the advantage that the MMG model is a modular-type model based on fluid dynamics. On the other hand, as the model becomes more sophisticated to accommodate low-speed maneuvering motions, the total number of parameters increases. Therefore, additional costs may be required, such as the maintenance of charts to determine those coefficients. The model builder decides whether to determine these coefficients from charts, tank tests, or SI from actual shipboard tests, based on the accuracy required and coding costs.

\subsection{Determine hydrodynamic derivatives by model experiment or model CFD computation}

When constructing a dynamic model using a captive model test or its CFD calculation results, the coefficients are obtained by fitting the measured fluid dynamic force with the dynamic model used in the simulation. In the following, the method of constructing a modular-type model is considered concerning the literature \cite{P29report_En}. Assuming that the resistance, self-propulsion performance, and propeller open performance of the subject vessel are known, the required tests are as follows.
\begin{enumerate}
    \item Steering test for different propeller load
    \item Steering test with drift and turning motions
    \item Drift motion test, turning motion test, and/or PMM test
\end{enumerate}
The steering tests are conducted to obtain coefficients related to the hull-rudder interaction and the longitudinal component of rudder inflow velocity in the propeller wake.
Drift tests, turning tests, and PMM tests all aim to acquire fluid dynamic forces acting on the hull. However, the acceleration component cannot be obtained from the drift tests and circular motion tests. Therefore, Motora charts \cite{Motora1959En,Motora1960En,Motora1960_2En} or theoretical calculations are usually used for obtaining the added mass component.
The PMM test enables model construction with fewer points than the drift and circular motion tests. In addition, added mass can be obtained by this test. On the other hand, it should be noted that the obtained fluid dynamic forces from the PMM test have a dependency on frequency.
In addition, by conducting the drift and circular motion tests with the propeller, it is possible to obtain the coefficients on the changes in the wake coefficient due to ship motion.
The steering test with drift and circular motion makes it possible to obtain coefficients related to the sway directional component of inflow velocity at the rudder position (flow straightening coefficient).

The test conditions for the captive model test should be such that the motion state that the simulation should cover and the operating state of the propeller and rudder are not extrapolated from the test conditions. In addition, the measurement points should be set at intervals that allow for good interpolation between them. For example, if the purpose of the test is to simulate normal maneuvers with normal ship velocity, the following test conditions can be considered:
\begin{itemize}
\item Steering test for different propeller loads is performed with a rudder angle of -35 to 35 deg. at 5 deg. intervals and three different propeller loads.
\item The drift test is performed with a drift angle of -20 to 20 deg. at 5 deg. intervals. the circular motion test is performed with a dimensionless yaw angular velocity of -0.8 to 0.8 at 0.2 intervals, with a drift angle as needed.
\item The steering test with drift and circular motion is performed for three different steering angles under the same conditions as the drift and circular motion test.
\end{itemize}

These are the test conditions for creating a dynamic model that can simulate maneuvers with normal maneuvers with normal ship velocity. On the other hand, when creating a dynamic model that can simulate maneuvers with normal maneuvers with normal ship velocity, it is necessary to conduct additional tests. For example, pivot-turn tests with a drift angle of 90deg. or more, propeller and rudder characteristics in each of the four quadrants of the propeller operational condition, and tests to measure shallow-water effects for all of them will be necessary. Thus, the number of tests will increase even more.

\subsection{System identification (SI) using direct maneuvering simulation or the trajectory from the full-scale/model-scale ship.}
The System Identification (SI) approach involves preparing time-series data of the vessel's trajectory, i.e., the time-dependent data of state variables, control inputs, and disturbances, as training data. The model's parameters (including its structure, if necessary) are then searched using some optimization method to match the training data, which may include fluid dynamic forces such as direct pressure on the rudder. Therefore, the cost associated with acquiring trajectory data becomes a part of the construction cost. Although the cost of obtaining trajectory data depends on the vessel's hourly charter cost, we will focus here solely on the dimensionless length of the learning data without considering charter costs.



The training data quantity for system identification in related studies is presented in \cref{tab:SI_data}. In the table, the length of the training data, $T^{(\text{train})}$ (sec), is non-dimensionalized using the ship's length $L$ (m) and representative velocity $V$ (m/s) of the target ship to facilitate the comparison of diverse studies considering variations in motion types and vessel sizes.
\begin{equation}
    T^{\prime(\text{train})} = T^{(\text{train})}V/L \enspace.
\end{equation}

\begin{table*}
    \centering
    \caption{Data set used in relevant SI studies. In the maneuver column, Z, T and R are a zig-zag maneuver, a turning and a random maneuver. See literature \cite{Miyauchi2022SI,Wakita2022}, v were set to $Fr = 0.05$.}
    \begin{tabular}{lp{8em}p{5em}p{5em}p{5em}}
        Article & Model  & Unknown parameters & Maneuver & $T^{\prime(\text{train})}$ \\
        \hline\hline
        \cite{Maki2014ASR_En}    & KT & 2&Z & 14 \\
        \cite{Abkowitz1980} & Whole-ship, linear &15 & Z& 17.7 \\
        \cite{Araki2012a}    & Modular-type (MMG) & 21, 30&Z, T & 8.8-50.8 \\
        \cite{Wang2020} & Whole-ship & 7, 18 & m-level PRS & 25.9 \\
        \cite{Miyauchi2022SI} & Modular-type (MMG) & 57 & Z, T, R & 258-269.5 \\
        \cite{Wakita2022} & RNN & 122600 & Z, T, R & 256.3  \\
        \hline
    \end{tabular}
    \label{tab:SI_data}
\end{table*}

In the case of using turning and Zigzag (Z) tests, the initial velocity of the ship was denoted by $V$. As a reference for the non-dimensionalized data quantity $T^{\prime(\text{train})}$, the data employed by Abkowitz et al. \cite{Abkowitz1980} consisted of a single trajectory of a 10-degree/10-degree Z-test up to the 4th overshoot with a ship length of $L=325~\mathrm{m}$ and an initial velocity of $V=7~\mathrm{knots}$. In this particular case, $T^{\prime(\text{train})}$ was equal to 17.7.
According to \cref{tab:SI_data}, even approaches aiming to estimate the model parameters of Whole-ship models or Modular-type models have successfully obtained dynamic model parameters from just one to a few Zigzag or turning test data. However, in the case of Whole-ship models, model parameters may change due to changes in the vessel speed. Therefore, obtaining dynamic models for several speeds would require several times more data. Additionally, for Modular-type models capable of representing changes in propeller revolutions per minute (RPM), models obtained solely from trajectory data with a single propeller RPM cannot capture the model parameters for any arbitrary propeller RPM, including changes in propeller rotation direction.

Therefore, some studies \cite{Araki2012a,Sutulo2014a} have sought model parameters related to thrust through methods other than System Identification (SI) and have exclusively used SI to estimate parameters related to ship fluid dynamic forces. To obtain all parameters by using SI, as in the case of Whole-ship models, a greater amount of training data would likely be necessary.


On the other hand, when using a dynamic model for a simulator, it is desirable to obtain a dynamic model that can respond to arbitrary inputs within the mechanically feasible range of the ship. Attempts to achieve a robust model for various inputs by using complex motions other than IMO Standard Maneuvers have been discussed since the early stages of System Identification (SI) research \cite{Abkowitz1980}.

In recent years, methods have been proposed to optimize the steering angle and steering time in maneuvers to improve the distribution of state variables, using motions such as steering maneuvers beyond the Zigzag test \cite{Yoon2003,Wang2020}. Miyauchi et al. \cite{Miyauchi2022SI} estimated an MMG model that can handle propeller reversal and large yaw angles by using random ship maneuvers as training data. Similarly, Wakita et al. \cite{Wakita2022} used random ship maneuvers to represent dynamic models with Recurrent Neural Networks (RNN). However, these random motions have drawbacks since a long measurement record is necessary to ensure sufficient randomness. Further, the real-scale random maneuvers may face the limit of the steering device's capability and safety issues during random maneuvers.

Additionally, when training a dynamic model using the free-running model experiments, it is crucial to be mindful of the unavoidable scale effects. Since Reynolds numbers differ between full-scale and model-scale ships, the frictional resistance coefficient and the wake coefficient also change. Although the contributions to each propeller load tend to cancel each other out, model-scale ships often experience higher propeller load.

To correct for the differences in propeller load, some studies \cite{Tsukada2014_En} have proposed methods involving the use of auxiliary thrust to perform corrections, such as adjusting the skin friction coefficient (SFC) during model tests. Another approach involves focusing on the inflow velocity to the rudder and making corrections based on this parameter \cite{Ueno2014_En}.



\section{Functional and non-functional requirements of the simulator for autonomous vessel}\label{sec:functional_requirement}
This chapter describes the functional requirements for the simulation of the motion of an automated ship. The requirements of the simulator should be defined according to its purpose.
\subsection{Example of functional requirements in IEC 62065}
Some existing simulators have been created with explicit functional requirements. For example, international standards such as IEC 62065 \cite{IEC62065} have already defined functional requirements for simulators of ship maneuvering motion.
IEC 62065 describes the following functional requirements for mathematical models that aim to provide a simple method for testing track control systems.

\begin{enumerate}
\item The model should be capable of representing the essential characteristics of ship motion such as:
    \begin{itemize}
        \item Response of propulsion system to command
        \item Response to propulsive force
        \item Straight line resistance
        \item Rudder response to a command
        \item Response of the hull to the actual steering angle
        \item Lateral motion of the hull when turning
    \end{itemize}
    \item The model should be able to represent the behavior of a vessel with unstable straight-line motion, i.e., it should produce a turning motion even when the rudder is neutral, and require rudder operation to suppress it.
    \item The model should be able to connect to commercial ship control systems without any special modifications. This means that inputs to the model should come from standard equipment and the outputs of the model should be suitable as direct inputs to standard equipment.
    \item The model should be simple enough that test houses around the world can interpret the equations without ambiguity and incorporate the formulas into computer simulators with minimum difficulty.
\end{enumerate}

\subsection{Proposal of functional and non-functional requirements}
For special maneuvers such as berthing/unberthing operations, the functional requirements of the IEC 62065 are considered to be insufficient.
Hereafter, the authors consider a model for maneuvering simulators that can be used for the development of control algorithms.
To achieve a model that is realistic and not overly complex, including berthing/unberthing maneuvering, the authors propose functional and non-functional requirements.
The control law to be developed will likely be a feedback control since it will be used under disturbance. The functional requirements listed below are for a simulator to develop a feedback controller to be installed on an automatic vessel. Feedback control is expected to compensate for modeling errors. Therefore, it is important to note that the simulation results give qualitatively plausible ship motion results. Therefore, the most important feature of the functional requirements presented here is that they do not require a high level of quantitative agreement. On the basis of the above, the functional requirements can be described in detail as follows.

\begin{itemize}
    \item Collision avoidance maneuver by heading control 
    \begin{enumerate}
        \item That the hull is subjected to hydrodynamic resistance in the surge, sway, and yaw directional hydrodynamic resistance acts on the hull due to the change in each velocity.
        \item Model tends to be stable or unstable depending on the hull form and velocity.
        \item Theoretically or physically improbable motion does not occur.
        \item According to the rudder angle and hull form, turning motion is generated.
        \item Ship velocities appropriately vary with propeller revolution number.
        \item The response of the ship generated by steering varies with ship velocity and propeller revolution.
        \item When wind and wave disturbance is present, the ship drifts to leeward or turns.
        \item The delay between control commands and the actuator is taken into account. 
        \item The maximum values of the actuator and its rate can be considered.
        \item The time steps of the differential equation are chosen automatically and appropriately according to the scale, etc.
        \item The calculation of each step must be completed in real-time.
    \end{enumerate}

    \item Berthing maneuver
    \begin{enumerate}
        \item In the absence of external forces and control inputs, etc., the system origin $(u,~v,~r)^{\top}=0$ has asymptotic stability.
        \item That the hull is subjected to hydrodynamic resistance in the surge, sway, and yaw directional hydrodynamic resistance acts on the hull due to the change in each velocity.
        \item Model tends to be stable or unstable depending on the hull form and velocity.
        \item Theoretically or physically improbable motion does not occur.
        \item According to the rudder angle and hull form, turning motion is generated.
        \item Ship velocities appropriately vary with propeller revolution number.
        \item The response of the ship generated by steering varies with ship velocity and propeller revolution.
        \item The response of the ship generated by side thrusters varies with the ship's forward velocity.
        \item In the case of propeller reversing, a turning motion is generated according to the direction of rotation of the individual ship's propeller.
        \item When wind disturbance is present, the ship drifts to leeward and/or turns.
        \item The delay between control commands and the actuator is taken into account. 
        \item The maximum values of the actuator and its rate can be considered
        \item The time steps of the differential equation are chosen automatically and appropriately according to the scale, etc.
        \item The calculation of each step must be completed in real-time.
    \end{enumerate}
\end{itemize}

On the other hand, as non-functional requirements, IEC 62065 states that "the model shall be publicly available" and "the model shall be as possible as simple". On the other hand, there may be no particular need for model simplification. In this research, the following are proposed as non-functional requirements.
\begin{enumerate}
    \item The model is open source
    \item The model is clearly described using mathematical equations.
\end{enumerate}

Based on the above, the user should design and use a model after defining the functional requirements of the model, i.e., what items should be represented. Even when selecting an existing model, the user should use one that meets the established functional requirements.

Finally, the authors show in the Appendix the maneuvering motion model that satisfies the functional requirements proposed here, and its code is available on \url{https://github.com/NAOE-5thLab/mmg-1p1r-simulator}.

\section{Notes on solving differential equations numerically}\label{sec:ODE_solve}
Suppose that the equations of motion can be described numerically according to the functional requirements described so far. Then the next step is to solve the differential equation numerically. 
Then, all of the equations involved must be coupled and the state quantities must be computed simultaneously for the control input $\boldsymbol{u}(t)$ and the disturbance $\boldsymbol{\omega}$ by \cref{eq:integral_recursive}. Note that this is a rewritten representation.
\begin{equation}
    \boldsymbol{x}_t = \boldsymbol{x}_0+\int_{0}^{t} \boldsymbol{F}\left(\boldsymbol{x}_{s}, \boldsymbol{u}\left(\boldsymbol{x}_{s}(\boldsymbol{\omega})\right)\right) \mathrm{d} s
    \label{eq:integral_recursive}
\end{equation}
In addition, it should be noted that \cref{eq:motion} must be symbolically solved for $\dot{u}$, $\dot{v}$, and $\dot{r}$. Then, the following form must be taken and then solved numerically.

\begin{equation}
    \left\{
    \begin{aligned}
        \dot{u}&=f_u(u,\,v,\,r)\\
        \dot{v}&=f_v(u,\,v,\,r)\\
        \dot{r}&=f_r(u,\,v,\,r)
    \end{aligned}
    \right.
\end{equation}

\section{Discussion}\label{sec:discussion}


In this chapter, building upon the discussions regarding the existing models, we delve into the challenges associated with the introduction of models into the simulator.
The process of designing a dynamic model for use in a simulator can be divided into several subtasks. These are listed and discussed below.
\begin{enumerate}
    \item Select a appropriate model for your purpose.
    \item Clearly identify the inputs, outputs, and parameters of the selected model.
    \item Design of input/output interface and protocols. 
    \item Design appropriate parameters of the model.
\end{enumerate}

\begin{figure}
    \centering
    \includegraphics[width=1.0\hsize]{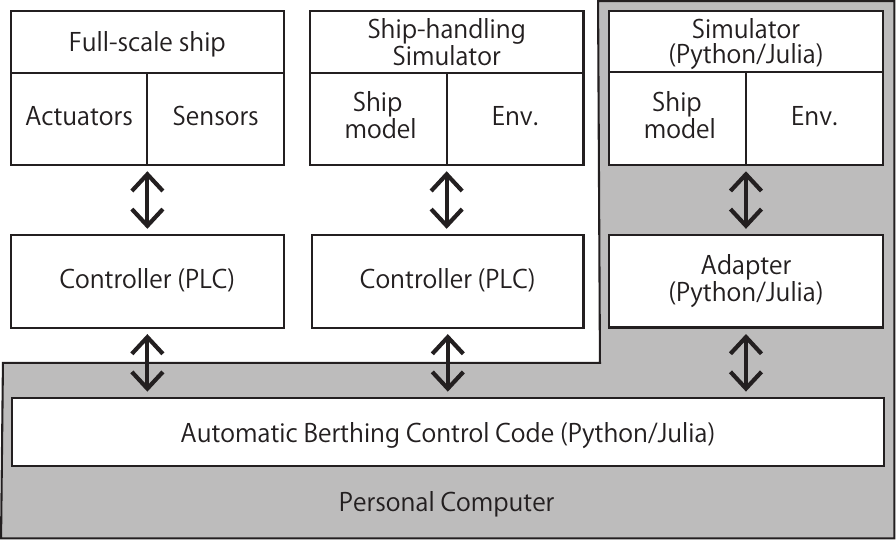}
    \caption{Development environment for prototyping of automatic berthing system in \cite{sawada_dissertation}}
    \label{fig:dev_env}
\end{figure}

How to select a model for the purpose is described in previous sections.
Before selecting a model, you need to design the requirements of the model for your purpose.
Identifying the inputs, outputs, and parameters of the selected model and designing interfaces and protocols are also important in developing automated systems.
The preceding discussions have primarily focused on the requirements of the model. 
However, it is imperative for the simulator to furnish an environment that ensures the quality of the developed automated system.
For example, Fig. \ref{fig:dev_env} shows the development environment of the automatic berthing system in the study by Sawada \cite{sawada_dissertation}.
When introducing the chosen model into the simulator for MASS under development, it is desirable that it can simulate the communication interface that the actual developed MASS has.
For instance, communication specifications with actuators on the real ship and data receiving frequencies of onboard sensors can impact control.
If the communication method connecting the models within the simulator and the developed control code or automated system closely resembles the specifications of the systems on the actual ship, it can reduce malfunctions attributed to the hardware of the automated system and onboard system.
This method is called HILS (Hardware In the Loop Simulation).
In this case, actuators and sensors of the full-scale ship are connected to the controller, and the developed control code communicates with the controller to operate the ship.
The ship-handling simulator is updated to develop the automated system and connected with the same controller as the one on the real ship.
The simulator on the personal computer which is used in the early stage of development is implemented in the same runtime environment as the control code.
This simulator has an adapter that simulates the communication between the control code and the onboard controller.
When implementing a dynamic model in a simulator, the dynamic model should be designed considering such a hardware interface.

Parameter design relates to the availability of the model.
Many papers discuss parameter estimation methods, however, there are no systematic and comprehensive studies or guidelines on this subject. In addition, when adopting MMG models, for example, costs of implementation are high, and source code and model implementation examples are not publicly available.
The process of selecting a model, collecting data, and implementing the model in a simulator is time-consuming and costly.
For such problems, it would be helpful for the development of an autonomous control system if examples of verified implementations of the model were made publicly available.

In the development of algorithms for automatic ship operations, especially in the development of a simulator including models to be applicable to berthing/unberthing situations, the MMG model may be selected if there is plenty of background information available. Otherwise, other models should be used.
In addition, it is necessary to standardize the structure of the model with an expressive power that can be applied to the simulation of a jetty at takeoff and landing, to develop a chart of the coefficients, and to actively open source the code so that ordinary users can also build the model. Once this is achieved, MMG models should be easily constructed by organizations that are not experts in fluid dynamics or maneuvering equations of motion, and that do not have tank testing facilities or background information.

In the development of algorithms for autonomous ship operation, the choice of model should be based on requirements. For example, when developing a simulator for berthing/unberthing situations, the MMG model, modified Abukowitz model, and Fossen's model are likely to meet the requirements for the situations.
However, as discussed so far, there is a trade-off between expressiveness and implementation cost. Implementation cost can be improved by utilizing the existing database of aquarium tests, but there is a problem of loss of parameter compatibility if the model differs.
Standardization of models is one solution, and it is important to maintain a database of parameters of standardized models and to actively open source the calculation codes of models.

Whereas, the use of response models, modified Abkowitz model, neural network-based model, or Fossen's model is always an option. 
The modified Abkowitz model and neural network-based model require training data.
In addition, a database of response model coefficients applicable to berthing operations has not been developed.

\section{Concluding remarks}\label{sec:conclusion}


In this paper, against the background of the widespread use of simulators in the development of automated vessels, the authors show the importance of organizing the functional requirements according to the evaluation, task, etc., and of appropriately selecting a dynamic model.
In the development of an automatic ship, it is important to discuss the cost and functionality of the dynamic model in addition to the hydrodynamic theory behind the model.

In this study, the authors have organized the applications of dynamic models in general, and in particular, the authors have systematically organized the typical dynamic models that are often used for self-propelled vessels. 
Furthermore, examples of functional and non-functional requirements for dynamic models in the development of automatic vessels are presented, taking the collision avoidance maneuver and berthing/unberthing maneuver as examples of tasks for autonomous ships. 
Further research will be conducted to ensure that dynamic models are appropriately selected according to the flow of identifying requirements as presented in this paper.

Finally, the authors proposed to standardize the model structure to facilitate the development of simulators so that dynamic models can be built by the general public.
The authors believe that it is necessary to standardize the model structure, develop charts for parameters of the dynamic models, and open-source the calculation code in order to promote the development of simulators so that general users can build dynamic models.


\begin{acknowledgements}
This study was supported by a Grant-in-Aid for
Scientific Research from the Japan Society for Promotion of Science (JSPS KAKENHI Grant \#22H01701)
\end{acknowledgements}

%
\section*{Conflict of interest}

    The authors declare that they have no conflicts of interest.

\appendix
\section*{Appendix}

In this Appendix, the authors present a maneuvering motion model that satisfies the functional requirements proposed in this paper for the berthing maneuver in which a ship is operated at low speed. The model is completely identical to that shown in Miyauchi et al. \cite{Miyauchi2022SI}. In his paper, the full coefficient list for this model is available. The Python implementation is available at \url{https://github.com/NAOE-5thLab/mmg-1p1r-simulator}.

The 3-DoF equation of MMG model \cite{Miyauchi2022SI} is express as follows:
\begin{equation}\label{eq:MMGdynamics}
    \begin{split}
    (m + m_x) \dot{u} - (m + m_y) v_{\mathrm{m}} r -x_{\mathrm{G}}mr^{2} &= X\\
    (m + m_y) \dot{v}_{\mathrm{m}} + (m + m_x) u r +x_{\mathrm{G}}m\dot{r} &= Y\\
    (I_{zz} + J_{zz}+x_{\mathrm{G}}^{2}m) \dot{r} + x_{\mathrm{G}} m (\dot{v}_{\mathrm{m}} + ur) &= N
    \end{split}
\end{equation}
with
\begin{equation}
    \begin{split}
    X &= X_{\mathrm{H}} + X_{\mathrm{p}} + X_{\mathrm{R}} + X_{\mathrm{wind}} \\
    Y &= Y_{\mathrm{H}} + Y_{\mathrm{p}} + Y_{\mathrm{R}}  + Y_{\mathrm{wind}}\\
    N &= N_{\mathrm{H}} + N_{\mathrm{p}} + N_{\mathrm{R}} + N_{\mathrm{wind}}\enspace.
    \label{eq:MMGcompose}
    \end{split}
\end{equation}
Here, the overdot (e.g. $\dot{x})$ denotes the differentiation with respect to time. The right-hand side of \Cref{eq:MMGdynamics} represents the force or moment acting on the ship, and the MMG model decomposes the fluid dynamic force acting on the ship into a sub-model for major components consisting of the ship as \Cref{eq:MMGcompose}. The subscripts H, P, R, and wind denote the hull, the propeller, the rudder, and the external forces by wind, respectively. Hereafter, the details of each sub-model are explained. 
\\
\begin{enumerate}
    \item \textbf{Force on Hull}
    \\
    The force acting on the hull was estimated by the following model for operation on the open sea and harbor maneuvers~\cite{Yoshimura2009a}:
    \begin{align}\label{eq:yoshimura}
        \begin{aligned}
            X_{\mathrm{H}}=&\left(\frac{\rho}{2}\right) L_{\mathrm{pp}} d
            \left[ \begin{array}{ll}\begin{split}
                &\left\{X_{0(\mathrm{F})}^{\prime}+\left(X_{0(\mathrm{A})}^{\prime}-X_{0(\mathrm{F})}^{\prime}\right)(\beta / \pi)\right\} u U \\
                &+X_{v r}^{\prime} L_{\mathrm{pp}} \cdot v_{\mathrm{m}} r
                \end{split}\end{array}\right] \\
            Y_{\mathrm{H}}=&\left(\frac{\rho}{2}\right) L_{\mathrm{pp}} d\\
            &\quad \left[ \begin{array}{ll}\begin{split}
                &Y_{v}^{\prime} v_{\mathrm{m}}|u|+Y_{r}^{\prime} L_{\mathrm{pp}} \cdot r u \\
                &-\left(\frac{\underline{C_{\mathrm{D}}}}{L_{\mathrm{pp}}}\right) \int_{-L_{\mathrm{pp}} / 2}^{L_{\mathrm{pp}} / 2}\left|v_{\mathrm{m}}+C_{rY} r x\right|\left(v_{\mathrm{m}}+C_{rY} r x\right) d x
            \end{split}\end{array}\right] \\
            N_{\mathrm{H}}=&\left(\frac{\rho}{2}\right) L_{\mathrm{pp}}^{2} d \\
            &\quad \left[ \begin{array}{ll}\begin{split}
                &N_{v}^{\prime} v_{\mathrm{m}} u+N_{r}^{\prime} L_{\mathrm{pp}} \cdot r|u| \\
                &-\left(\frac{\underline{C_{\mathrm{D}}}}{L_{\mathrm{pp}}^{2}}\right) \int_{-L_{\mathrm{pp}} / 2}^{L_{\mathrm{pp}} / 2} \left|v_{\mathrm{m}}+C_{rN} r x \right| \left(v_{\mathrm{m}}+C_{r N} r x\right) x d x
            \end{split}\end{array}\right]\enspace,
        \end{aligned}
    \end{align}
    where $\rho,$ density of water; $L_{\mathrm{pp}}$, Length between perpendiculars of ship; $d$, draft of ship; $X_{O_(\mathrm{F})}^{\prime} \text{and} \ X_{0(\mathrm{A})}^{\prime}$, non-dimensional resistance coefficients of ahead and astern; $C_{\mathrm{D}}$, cross flow drag coefficient; $C_{r Y} \ \text{and } C_{r N}$, correction factor for lateral force and yaw moment; $X_{0(\mathrm{F})}^{\prime}, \ Y_{v}^{\prime}, \ Y_{r}^{\prime}, \ N_{v}^{\prime}, \ \text{and}\ N_{r}^{\prime}$ are non-demensional hydrodynamic derivatives, respectively. Hereafter, the superscript prime (e.g.\ $ X_{0(\mathrm{F})}^{\prime})$ means the non-dimensionalized value and the non-denationalization is conducted by \Cref{eq:nondimentionalization}. Note that in this paper, the added mass components in the original expression of \Cref{eq:yoshimura} was moved to the left- hand side of \Cref{eq:MMGdynamics}.
    \\
    \item \textbf{Force by propeller}
    \\
    In the standard MMG model~\cite{Yasukawa2015MMG}, the forwarding maneuver $(u>0,\ n_{\mathrm{p}}>0)$ is the target. In order to consider every propeller operation condition, the following quadrant operational condition is to be taken into account: first $(u\geq0,\ n_{\mathrm{p}}>0)$; second $(u<0,\ n_{\mathrm{p}}>0)$; third $(u\geq0,\ n_{\mathrm{p}}<0)$; and fourth $(u<0,\ n_{\mathrm{p}}<0)$. On the first and second quadrants, propeller thrust was estimaed by the standard MMG model:
    \begin{equation}
        X_{\mathrm{p}}=\rho n_{\mathrm{p}}^{2} D_{\mathrm{p}}^{4}\left(1-t_{\mathrm{p}}\right) K{\mathrm{T}}\enspace,
    \end{equation}
    where thrust coefficient $K{\mathrm{T}}$ was express by a polynomial expression of advance coefficient $J_{\mathrm{p}}=\left(1-w_{\mathrm{p}}\right) u /\left(n_{\mathrm{p}} D_{\mathrm{p}}\right)$.
    The effective propeller wake fraction $w_{\mathrm{p}}$ was calculated as follows~\cite{ITTC2002}:
    \begin{equation}
        1-w_{\mathrm{p}}=1-w_{\mathrm{p} 0} + \tau\left|v_{\mathrm{m}}^{\prime}+x_{\mathrm{p}}^{\prime} r^{\prime}\right|+C_{\mathrm{p}}^{\prime}\left(v_{\mathrm{m}}^{\prime}+x_{\mathrm{p}}^{\prime} r^{\prime}\right)^{2} \enspace,
    \end{equation}
    where: $w_{\mathrm{p0}}$ is the wake fraction on $v_{\mathrm{m}}=r=0$; $\tau, \ C^{\prime}_{\mathrm{p}} \text{and} \ x^{\prime}_{\mathrm{p}}$ are empirical coefficients. The trust deduction factor $t_p$ and wake fraction $w_{\mathrm{p}}$ changes due to propeller operation condition. However, in our model, they are simply modeled as ~\cite{KOBAYASHI1994,Yasukawa2003}.:
    \begin{align}
        t_{\mathrm{p}}=0 \quad &\text{for } n_{\mathrm{p}}<0\\
        w_{\mathrm{p}}=0 \quad &\text{for } u<0 \enspace,
    \end{align}
    Sway force and yaw moment induced by the propeller on the first and second quadrants $(n_{\mathrm{p}}\geq 0)$ are usually neglected in the MMG model. However, in our model, they were estimated by the polynomial expression based upon the captive test for training vessel~\cite{Ueno2001}, as follows:
    \begin{align}
        Y_{\mathrm{p}} = \begin{cases}
        0 &\text{for } u\geq0\\
        \frac{1}{2} \rho L_{\mathrm{pp}}^{2} d \left(n_{\mathrm{p}} P\right)^{2}\left(A_{6} J_{\mathrm{s}}^{2}+A_{7} J_{\mathrm{s}}+A_{8}\right) &\text{for } u<0
        \end{cases}\\
        N_{\mathrm{p}} = \begin{cases}
        0 &\text{for } u\geq0\\
        \frac{1}{2} \rho L_{\mathrm{pp}}^{2} d \left(n_{\mathrm{p}} P\right)^{2}\left(B_{6} J_{\mathrm{s}}^{2}+B_{7} J_{\mathrm{s}}+B_{8}\right) &\text{for } u<0
        \end{cases}
    \end{align}
    where: $P$, pitch of the propeller; $J_{\mathrm{s}}=u/(n_{\mathrm{P}}D_{\mathrm{P}})$; $A_{6} \ \text{through} \ A_{8}$ and $B_{6} \ \text{through} \ B_{8}$ are polynomial coefficients.
    
    On the propeller reversal condition, same as the second quadrant, polynomial expression~\cite{Hasegawa1994} based on CMT was used:
    \begin{align}
        X_{\mathrm{p}} =& \rho n_{\mathrm{p}}^{2} D_{\mathrm{p}}^{4}\begin{cases}
            C_{6}+C_{7} J_{\mathrm{s}} & \text{for~} \left(J_{\mathrm{s}} \geq C_{10}\right) \\
            C_{3} & \text{for~} \left(J_{\mathrm{s}}<C_{10}\right)
        \end{cases} \\
        Y_{\mathrm{p}}=&\frac{1}{2} \rho L d\left(n_{\mathrm{p}} D_{\mathrm{p}}\right)^{2}
        \begin{cases}
            A_{1}+A_{2} J_{\mathrm{s}} & \left(-0.35 \leq J_{\mathrm{s}} \leq -0.06\right) \\
            A_{3}+A_{4} J_{\mathrm{s}} & \left(J_{\mathrm{s}}<-0.35\right) \\
            A_{5} & \left(-0.06<J_{\mathrm{s}}\right)
        \end{cases}\\
        N_{\mathrm{p}}=&\frac{1}{2} \rho L^{2} d\left(n_{\mathrm{p}} D_{\mathrm{p}}\right)^{2}\begin{cases}
            B_{1}+B_{2} J_{\mathrm{s}} & \left(-0.35 \leq J_{\mathrm{s}} \leq -0.06\right) \\
            B_{3}+B_{4} J_{\mathrm{s}} & \left(J_{\mathrm{s}}<-0.35\right) \\
            B_{5} & \left(-0.06<J_{\mathrm{s}}\right) \enspace, 
        \end{cases}
    \end{align}
    where, $A_{1} \ \text{through} \ A_{5}$, $B_{1} \ \text{through} \ B_{5}$, $C_{3}, \ C_{6}, \ C_{7},$ and $C_{10}$ are polynomial coefficients.
    \\
    \item \textbf{Force by Rudder}
    \\
    The forces and moment induced by the rudder is estimated by the standard MMG model~\cite{Yasukawa2015MMG}:
    \begin{align}
        X_{\mathrm{R}}&=-\left(1-t_{\mathrm{R}}\right) F_{N} \sin \delta \\
        Y_{\mathrm{R}}&=-\left(1-a_{\mathrm{H}}\right) F_{N} \cos \delta \\
        N_{\mathrm{R}}&=-\left(x_{\mathrm{R}}+a_{\mathrm{H}}\ x_{\mathrm{H}}\right) F_{N} \cos \delta \enspace, 
    \end{align}
    where $F_{N}$ is the rudder normal force:
    \begin{equation}
        F_{N}=(1 / 2) \rho A_{\mathrm{R}} U_{\mathrm{R}}^{2} f_{\alpha} \sin \alpha_{\mathrm{R}}\enspace.
    \end{equation}
    Here, $t_{\mathrm{R}}$, thrust deduction factor by steering; $x_{\mathrm{R}}$, longitudinal position of the rudder from midship; $a_{\mathrm{H}}$, correction factor lateral force; $x_{\mathrm{H}}$ position of additional lateral force on hull due to steering; $A_{\mathrm{R}}$, area of the rudder. The gradient of normal force of the rudder $f_{\alpha}$ is regarded as constant in most of the MMG model. Although, $f_{\alpha}$ is no longer constant in the case of the berthing maneuver, $f_{\alpha}$ is assumed as constant and estimated by the most commonly used Fujii's formula~\cite{Fujii_rudder_coef}. This formula has a function form with respect to the aspect ratio of rudder $\lambda$:
    \begin{equation}
        f_{\alpha} = 6.13\lambda/(2.25+\lambda)\enspace.
    \end{equation}
    The rudder inflow velocity $U_{\mathrm{R}}$ and effective inflow angle $\alpha_{\mathrm{R}}$ were estimted by surge and sway directional inflow velocities $u_{\mathrm{R}}$ and $v_{\mathrm{R}}$:
    \begin{align}
        U_{\mathrm{R}}&=\sqrt{u_{\mathrm{R}}^{2}+v_{\mathrm{R}}^{2}} \\
        \alpha_{\mathrm{R}}&=\delta-\mathtt{atan2}\left(\frac{v_{\mathrm{R}}}{u_{\mathrm{R}}}\right) \enspace.\label{eq:atan2}
    \end{align}
    In order to simulate the berthing maneuver, we extended the standard MMG model to \Cref{eq:atan2} by introducing the function $\mathtt{atan2}(y/x)$. Here, $\mathtt{atan2}(y/x)$ returns the $\tan^{-1}(y/x)$ in range of $(-\pi, \pi]$. The sway directional inflow velocity $v_{\mathrm{R}}$ is estimated as follows using the flow straightening coefficient $\gamma$ and experimental constant $l_{\mathrm{R}}$:
    \begin{equation}
    v_{\mathrm{R}}=\begin{cases}
        -\gamma_{\mathrm{p}}\left(v_{\mathrm{m}}+l_{\mathrm{R}} r\right) & \text{for } v_{\mathrm{m}}+x_{\mathrm{R}} r \geq 0\ \\
        -\gamma_{N}\left(v_{\mathrm{m}}+l_{\mathrm{R}} r\right) & \text{for } v_{\mathrm{m}}+x_{\mathrm{R}} r < 0 \enspace .
    \end{cases}
    \end{equation}
    The surge directional inflow $u_{\mathrm{R}}$ is strongly affected by the ship motion and the propeller-induced velocity. For $n_{\mathrm{p}}\geq0$, $u_{\mathrm{R}}$ is estimated in modified form~\cite{Yoshimura1978} for low speed region as:
    \begin{equation}
    \begin{split}
        &u_{\mathrm{R}}=\\
        &\varepsilon \sqrt{\eta\left\{u_{\mathrm{p}}+\frac{k_{x}}{\varepsilon}\left(\sqrt{u_{\mathrm{p}}^{2}+\frac{8 K{\mathrm{T}}\left(n_{\mathrm{p}} D_{\mathrm{p}}\right)^{2}}{\pi}}-u_{\mathrm{p}}\right)\right\}^{2}+(1-\eta) u_{\mathrm{p}}^{2}} \enspace.
    \end{split}
    \end{equation}
    Here, $u_{\mathrm{p}}=(1-w_{\mathrm{p}})u$; $\eta=D_{\mathrm{P}}/H_{\mathrm{R}}$; $H_{\mathrm{R}}$, height of the rudder; $\varepsilon$, ratio of wake fraction; $k_{x}$, empirical coefficient. On the third quadrant, Kitagawa's model~\cite{Kitagawa2015} is applied:
    \begin{equation}
    u_{\mathrm{R}}=\operatorname{sgn}\left(u_{\mathrm{Rsq}}\right) \cdot \sqrt{\left|u_{\mathrm{Rsq}}\right|} \enspace,
    \end{equation}
    where:
    \begin{align}
     \begin{split}
       u_{\mathrm{Rsq}}=&\eta \cdot \operatorname{sgn}\left(u_{\mathrm{RPR1}}\right) \cdot u_{\mathrm{RPR1}}^{2}\\
       &\quad +(1-\eta) \operatorname{sgn}\left(u_{\mathrm{RPR2}}\right) \cdot u_{\mathrm{RPR2}}^{2}+C_{\mathrm{PR}} \cdot u
    \end{split}\\
        u_{\mathrm{RPR1}}&=u \varepsilon\left(1-w_{\mathrm{p}}\right)+n_{\mathrm{p}} D_{\mathrm{p}} k_{x PR} \sqrt{8\left|K{\mathrm{T}}\right| / \pi} \\
        u_{\mathrm{RPR2}}&=u \varepsilon\left(1-w_{\mathrm{p}}\right) \enspace.
    \end{align}
    Here, $k_{x \mathrm{PR}}$, the velocity increase factor; $C_{\mathrm{PR}}$, the correction factor for the propeller reversal condition.
    On the fourth quadrant, it is assumed that the inflow velocity is equal to the velocity of ship itself: $u_{\mathrm{R}}=u$~\cite{KOBAYASHI1994}.
    \\
    \item \textbf{Force by Wind}
    \\
    The wind-induced forces and moments estimated by Fujiwara's regression formulae~\cite{Fujiwara1998_En}:
    \begin{align}
        \begin{aligned}
            X_{\mathrm{wind}} &= (1/2)\rho_{\mathrm{A}}U_{\mathrm{A}}^{2}A{_{\mathrm{T}}}\cdot C_{X} \\
            Y_{\mathrm{wind}} &= (1/2)\rho_{\mathrm{A}}U_{\mathrm{A}}^{2}A_{\mathrm{L}}\cdot C_{Y} \\
            N_{\mathrm{wind}} &= (1/2)\rho_{\mathrm{A}}U_{\mathrm{A}}^{2}A_{\mathrm{L}}L_{\mathrm{OA}}\cdot C_{N} \enspace,
        \end{aligned}
    \end{align}
    where
    \begin{align}
        \begin{aligned}
            C_{X} =& X_{0}+X_{1} \cos (2\pi - \gamma_{\mathrm{A}})+X_{3} \cos 3 (2\pi - \gamma_{\mathrm{A}}) \\
                  &+X_{5} \cos 5 (2\pi - \gamma_{\mathrm{A}}) \\
            C_{Y} =& Y_{1} \sin (2\pi - \gamma_{\mathrm{A}})+Y_{3} \sin 3 (2\pi - \gamma_{\mathrm{A}}) \\
                  &+Y_{5} \sin 5 (2\pi - \gamma_{\mathrm{A}}) \\
            C_{N} =& N_{1} \sin (2\pi - \gamma_{\mathrm{A}})+N_{2} \sin 2 (2\pi - \gamma_{\mathrm{A}}) \\
                  &+N_{3} \sin 3 (2\pi - \gamma_{\mathrm{A}}) \enspace.
        \end{aligned}
    \end{align}
    Here, $\rho_{\mathrm{A}}$ is the density of air, $A_{\mathrm{T}},\ A_{\mathrm{L}}, L_{\mathrm{OA}}$ are the projected transverse area, the projected lateral area and the overall length of the ship, respectively. $X_{i}, \ Y_{i}, N_{i}$ are coefficients to express wind pressure coefficients derived from regression formulae~\cite{Fujiwara1998_En} that use the geometric parameters of the ship as explanatory variables and are based on wind tunnel test data from numerous scaled model ships.
    
\end{enumerate}

\bibliographystyle{spphys}       
\bibliography{main.bib}   

\begin{thebibliography}{100}
\providecommand{\url}[1]{{#1}}
\providecommand{\urlprefix}{URL }
\expandafter\ifx\csname urlstyle\endcsname\relax
  \providecommand{\doi}[1]{DOI \discretionary{}{}{}#1}\else
  \providecommand{\doi}{DOI \discretionary{}{}{}\begingroup \urlstyle{rm}\Url}\fi

\bibitem{ABS_guide}
{American Bureau of shipping (ABS)}.
\newblock Guide for autonomous and remote control functions (2021)

\bibitem{NK_guide}
{Class NK}.
\newblock Guidelines for automated / autonomous operation on ships (ver.1.0) (2020)

\bibitem{BV_guide}
{Bureau Veitas (BV)}.
\newblock Guidance note {NI 641 DT R01 E}, guidelines for automated / autonomous operation on ships (2019)

\bibitem{dnv_guide}
{DNV-GL}.
\newblock {DNVGL-CG-0264} autonomous and remotely operated ships (2018)

\bibitem{dnv_sim}
{DNV-GL}.
\newblock {DNVGL-ST-0033} maritime simulator systemsn (2023)

\bibitem{osp_PEDERSEN2020104799}
T.A. Pedersen, J.A. Glomsrud, E.L. Ruud, A.~Simonsen, J.~Sandrib, B.O.H. Eriksen, Towards simulation-based verification of autonomous navigation systems, Safety Science \textbf{129}, 104799 (2020)

\bibitem{Minami_2022}
M.~Minami, M.~Kobayashi, K.~Hikida, K.~Kokubun, Development of the comprehensive simulation system for autonomous ships, Journal of Physics: Conference Series \textbf{2311}(1), 012012 (2022)

\bibitem{Miyauchi2023Sim_JASNAOE}
Y.~Miyauchi, S.~Wada, S.~Hamada, K.~Wakita, R.~Sawada, T.~Taniguchi, H.~Koike, A.~Maki, 2022a-os1-6 perspective on the simulator for autonomous vessel development, Conference Proceedings The Japan Society of Naval Architects and Ocean Engineers \textbf{35}, 37 (2022)

\bibitem{Fujino1983_En}
M.~Fujino, Application of maneuverability research to design (in japanese), Proceedings of the 12th marine dynamics symposium (154), 295 (1983)

\bibitem{Shouji1989_En}
K.~Shouji, T.~Ishiguro, A study on the mathematical model for the maneuverability of ship in harbor, Journal of the Kansai Society of Naval Architects, Japan \textbf{212}, 123 (1989)

\bibitem{ITTC2021manuvalidation}
{Manoeuvring Committe of the 29th ITTC}, {Validation of Manoeuvring Simulation Models}.
\newblock in \emph{{ITTC Recommended Procedures and Guidelines}} (2021), pp. 1--14, 7.5-02-06-03

\bibitem{fossen2021handbook}
T.I. Fossen, \emph{Handbook of marine craft hydrodynamics and motion control}, 2nd edn. (John Wiley \& Sons, 2021)

\bibitem{Hara1981_En}
K.~Hara, H.~Kobayashi, K.~Nomoto, Shiphandling simulators and their applications pp. 213--242 (1981)

\bibitem{Oliveira2022}
R.P. de~Oliveira, G.~Carim~Junior, B.~Pereira, D.~Hunter, J.~Drummond, M.~Andre, Systematic literature review on the fidelity of maritime simulator training, Education Sciences \textbf{12}(11) (2022)

\bibitem{Sawada2021a}
R.~Sawada, K.~Sato, T.~Majima, Automatic ship collision avoidance using deep reinforcement learning with {LSTM} in continuous action spaces, Journal of Marine Science and Technology \textbf{26}, 509 (2020)

\bibitem{maki2020_1}
A.~Maki, N.~Sakamoto, Y.~Akimoto, H.~Nishikawa, N.~Umeda, Application of optimal control theory based on the evolution strategy (cma-es) to automatic berthing, Journal of Marine Science and Technology \textbf{25}(1), 221 (2020)

\bibitem{maki2021_1}
A.~Maki, Y.~Akimoto, U.~Naoya, Application of optimal control theory based on the evolution strategy (cma-es) to automatic berthing (part: 2), Journal of Marine Science and Technology \textbf{26}(3), 835 (2021)

\bibitem{Sawada2021b}
R.~Sawada, K.~Hirata, Y.~Kitagawa, E.~Saito, M.~Ueno, K.~Tanizawa, J.~Fukuto, Path following algorithm application to automatic berthing control, Journal of Marine Science and Technology \textbf{26}, 541 (2021)

\bibitem{Sawada2023}
Y.K. Ryohei~Sawada, Koichi~Hirata, Automatic berthing control under wind disturbances and its implementation in an embedded system., Journal of Marine Science and Technology \textbf{28}, 452 (2023)

\bibitem{Miyauchi2021PP}
Y.~Miyauchi, R.~Sawada, Y.~Akimoto, N.~Umeda, A.~Maki, Optimization on planning of trajectory and control of autonomous berthing and unberthing for the realistic port geometry, Ocean Engineering \textbf{245}, 110390 (2021)

\bibitem{RACHMAN2022warm}
D.M. Rachman, A.~Maki, Y.~Miyauchi, N.~Umeda, Warm-started semionline trajectory planner for ship’s automatic docking (berthing), Ocean Engineering \textbf{252}, 111127 (2022)

\bibitem{Wakita2022RL}
K.~Wakita, Y.~Akimoto, D.M. Rachman, Y.~Miyauchi, N.~Umeda, A.~Maki.
\newblock {Collision-probability reduction method of tracking control for automatic docking/berthing using reinforcement learning} (2022).
\newblock Manuscript submitted for publication

\bibitem{Suyama2022}
R.~Suyama, Y.~Miyauchi, A.~Maki, Ship trajectory planning method for reproducing human operation at ports, Ocean Engineering  (under review)

\bibitem{Bingham2019}
B.~Bingham, C.~Agüero, M.~McCarrin, J.~Klamo, J.~Malia, K.~Allen, T.~Lum, M.~Rawson, R.~Waqar, Toward maritime robotic simulation in gazebo.
\newblock in \emph{OCEANS 2019 MTS/IEEE SEATTLE} (2019), pp. 1--10

\bibitem{Sarda2016}
E.I. Sarda, H.~Qu, I.R. Bertaska, K.D. {von Ellenrieder}, Station-keeping control of an unmanned surface vehicle exposed to current and wind disturbances, Ocean Engineering \textbf{127}, 305 (2016)

\bibitem{fossen1996identification}
T.I. Fossen, S.I. Sagatun, A.J. S{\o}rensen, Identification of dynamically positioned ships, Control Engineering Practice \textbf{4}(3), 369 (1996)

\bibitem{fossen_2022}
T.I. Fossen, Line-of-sight path-following control utilizing an extended kalman filter for estimation of speed and course over ground from gnss positions, Journal of Marine Science and Technology \textbf{27}(1), 806 (2022)

\bibitem{Fossen1999tutorial}
T.I. Fossen, J.P. Strand, Tutorial on nonlinear backstepping: Applications to ship control  (1999)

\bibitem{Hamamatsu2008En}
M.~Hamamatsu, H.~Kagaya, Y.~Kohno, Application of nonlinear receding horizon control for ship maneuvering, Transactions of the Society of Instrument and Control Engineers \textbf{44}(8), 685 (2008)

\bibitem{OHTSUKA2004}
T.~Ohtsuka, {A continuation/GMRES method for fast computation of nonlinear receding horizon control}, Automatica \textbf{40}(4), 563 (2004)

\bibitem{Ioki2022_En}
T.~Ioki, S.~Miyoshi, Development of maneuvering system for realizing autonomous ships, Conference Proceedings The Japan Society of Naval Architects and Ocean Engineers \textbf{34}, 51 (2022)

\bibitem{Habu2022_En}
I.~Habu, S.~Miyoshi, Development of automatic berthing control system for realizing fully autonomous ships, Conference Proceedings The Japan Society of Naval Architects and Ocean Engineers \textbf{34}, 21 (2022)

\bibitem{RACHMAN2022DPS}
D.M. Rachman, Y.~Aoki, Y.~Miyauchi, N.~Umeda, A.~Maki.
\newblock {Experimental Low-speed Positioning System With VecTwin Rudder for Automatic Docking (Berthing)} (2022).
\newblock Manuscript submitted for publication

\bibitem{Akazaki1937En}
S.~Akazaki, Experimental investigations on the turning of ships, Journal of Zosen Kiokai \textbf{1937}(61), 369 (1937)

\bibitem{Akazaki1968En}
S.~Akazaki, On the pivoting point and the turning quality of ships, Journal of the Society of Naval Architects of Japan \textbf{1968}(124), 105 (1968)

\bibitem{davidson1946}
K.S. Davidson, L.~Schiff, Turning and course keeping qualities of ships, Transactions SNAME  (1946)

\bibitem{Motora1959En}
S.~Motora, On the measurement of added mass and added moment of inertia for ship motions, Journal of Zosen Kiokai \textbf{1959}(105), 83 (1959)

\bibitem{Motora1955En}
S.~Motora, Course stability of ships, Journal of Zosen Kiokai \textbf{1955}(77), 69 (1955)

\bibitem{Inoue1979En}
S.~INOUE, M.~HIRANO, Y.~HIRAKAWA, K.~MUKAI, The hydrodynamic derivatives on ship maneuverability in even keel condition, TRANSACTIONS OF THE WEST-JAPAN SOCIETY OF NAVAL ARCHITECTS \textbf{57}, 13 (1979)

\bibitem{Nomoto1956_En}
K.~Nomoto, K.~Taguchi, K.~Honda, S.~Hirano, On the steering qualities of ships, Journal of Zosen Kiokai \textbf{1956}(99), 75 (1956)

\bibitem{Abkowitz1964}
M.A. Abkowitz, {Lectures on ship hydrodynamics--Steering and manoeuvrability}.
\newblock Tech. rep., Hydro and Aerodynamic Laboratory, Lyngby, Denmark (1964)

\bibitem{Ogawa1977MMG_En}
A.~Ogawa, T.~Koyama, K.~Kijima, {MMG Report-I}, Bulletin of the Society of Naval Architects of Japan \textbf{575}, 192 (1977)

\bibitem{Yasukawa2015MMG}
H.~Yasukawa, Y.~Yoshimura, {Introduction of MMG standard method for ship maneuvering predictions}, Journal of Marine Science and Technology (Japan) \textbf{20}(1), 37 (2015)

\bibitem{Nonaka1990En}
K.~Nonaka, On the manoeuvring motion of a ship in waves, Transactions of the west-Japan society of naval architectures \textbf{80}, 73 (1990)

\bibitem{Yasukawa2006wave1En}
H.~Yasukawa, Simulations of ship maneuvering in waves, Journal of the Japan Society of Naval Architects and Ocean Engineers \textbf{4}, 127 (2006)

\bibitem{Yasukawa2006wave2En}
H.~Yasukawa, Simulations of ship maneuvering in waves, Journal of the Japan Society of Naval Architects and Ocean Engineers \textbf{7}(0), 163 (2008)

\bibitem{Eda1980rolling}
H.~Eda, Rolling and steering performance of high speed ships.
\newblock in \emph{13th Symposium on Naval Hydrodynamics} (1980), pp. 427--439

\bibitem{Son_Nomoto1981}
K.~Son, K.~Nomoto, On the coupled motion of steering and rolling of a high speed container ship, Journal of the Society of Naval Architects of Japan \textbf{1981}(150), 232 (1981)

\bibitem{Yasukawa2010RollEn}
H.~YASUKAWA, A consideration of roll-coupling effect on ship maneuverability, The Journal of Japan Institute of Navigation \textbf{123}, 153 (2010)

\bibitem{Nakato1981sympo_En}
M.~Nakato, Trends in ship maneuverability research pp. 1--8 (1981)

\bibitem{Nomoto1957En}
K.~Nomoto, T.~Taguchi, K.~Honda, S.~Hirano, On the steering qualities of ships, International Shipbuilding Progress \textbf{4}(35), 354 (1957)

\bibitem{Okamoto1972En}
H.~Okamoto, H.~Tamai, H.~Oniki, Correlation studies of manoeuvrability of full ships, Journal of the Society of Naval Architects of Japan \textbf{1972}(131), 189 (1972)

\bibitem{Nomoto1969En}
K.~Nomoto, K.~Karasuno, A new procedure of manoeuvering model experiment, Journal of the Society of Naval Architects of Japan \textbf{1969}(126), 131 (1969)

\bibitem{Kim1978En}
K.J. Sam, E.~Kobayashi, K.~Nomoto, Least mean square iteration analysis on zig-zag manoeuvre tests, Journal of the Society of Naval Architects of Japan \textbf{1978}(144), 40 (1978)

\bibitem{astrom_kallstrom_1976}
K.~Åström, C.~Källström, Identification of ship steering dynamics, Automatica \textbf{12}(1), 9 (1976)

\bibitem{kallstrom1979adaptive}
C.~Källström, K.~Åström, N.~Thorell, J.~Eriksson, L.~Sten, Adaptive autopilots for tankers, Automatica \textbf{15}(3), 241 (1979)

\bibitem{ohtsu_1980}
K.~Ohtsu, A statistical analysis of the characteristics of the ships' motions and its control, JOURNAL OF THE MARINE ENGINEERING SOCIETY IN JAPAN \textbf{15}(9), 752 (1980)

\bibitem{Iseki1998iir_En}
T.~Iseki, K.~Ohtsu, On-line identification of ship maneuverability indices by using iir filters, Journal of the Society of Naval Architects of Japan \textbf{1998}(184), 167 (1998)

\bibitem{JIANG2020107202}
H.~Jiang, S.~Duan, L.~Huang, Y.~Han, H.~Yang, Q.~Ma, Scale effects in ar model real-time ship motion prediction, Ocean Engineering \textbf{203}, 107202 (2020)

\bibitem{MOREIRA2003}
L.~Moreira, C.~{Guedes Soares}, Dynamic model of manoeuvrability using recursive neural networks, Ocean Engineering \textbf{30}(13), 1669 (2003)

\bibitem{Moreira2012}
L.~Moreira, C.G. Soares, Recursive neural network model of catamaran manoeuvring, International Journal of Maritime Engineering \textbf{154} (2012)

\bibitem{RAJESH2008}
G.~Rajesh, S.~Bhattacharyya, System identification for nonlinear maneuvering of large tankers using artificial neural network, Applied Ocean Research \textbf{30}(4), 256 (2008)

\bibitem{Zhang2013}
X.G. Zhang, Z.J. Zou, Black-box modeling of ship manoeuvring motion based on feed-forward neural network with chebyshev orthogonal basis function, Journal of Marine Science and Technology \textbf{18}, 42 (2013)

\bibitem{OSKIN2013}
D.A. Oskin, A.A. Dyda, V.E. Markin, {Neural Network Identification of Marine Ship Dynamics}, IFAC Proceedings Volumes \textbf{46}(33), 191 (2013)

\bibitem{Wakita2022}
K.~Wakita, A.~Maki, N.~Umeda, Y.~Miyauchi, T.~Shimoji, D.M. Rachman, Y.~Akimoto, On neural network identification for low-speed ship maneuvering model, Journal of Marine Science and Technology \textbf{27}, 772 (2022)

\bibitem{WOO2018}
J.~Woo, J.~Park, C.~Yu, N.~Kim, Dynamic model identification of unmanned surface vehicles using deep learning network, Applied Ocean Research \textbf{78}, 123 (2018)

\bibitem{Jiang2022}
Y.~Jiang, X.R. Hou, X.G. Wang, Z.H. Wang, Z.L. Yang, Z.J. Zou, Identification modeling and prediction of ship maneuvering motion based on lstm deep neural network, Journal of Marine Science and Technology \textbf{27}, 125 (2022)

\bibitem{Chislett1965}
M.S. Chislett, J.~Strom-Tejsen, Planar motion mechanism tests and full-scale steering and manoeuvring predictions for a mariner class vessel, International Shipbuilding Progress \textbf{12}, 201 (1965)

\bibitem{Abkowitz1980}
M.A. Abkowitz.
\newblock Measurement of hydrodynamic characteristics from ship maneuvering trials by system identification (1980)

\bibitem{Ogawa1981sympo_En}
A.~Ogawa, M.~Hamamoto, in \emph{Proceedings of the 3rd symposium on ship maneuverability} (Society of Naval Architects of Japan, 1981), chap.~II, pp. 9--25

\bibitem{P29report_En}
{Research Committee on Standardisation of Ship Manoeuvrability Prediction Models}.
\newblock report (2012).
\newblock \urlprefix\url{https://www.jasnaoe.or.jp/research/dl/report\_p-29.pdf}

\bibitem{Lamb1932}
H.~Lamb, \emph{Hydrodynamics} (Cambridge University Press, 1932)

\bibitem{Miyauchi2022SI}
Y.~Miyauchi, A.~Maki, N.~Umeda, D.M. Rachman, Y.~Akimoto, System parameter exploration of ship maneuvering model for automatic docking/berthing using cma-es, Journal of Marine Science and Technology  (2022)

\bibitem{Yoshimura2009a}
Y.~Yoshimura, I.~Nakao, A.~Ishibashi, {Unified Mathematical Model for Ocean and Harbour Manoeuvring}.
\newblock in \emph{Proceedings of MARSIM2009} (2009), pp. 116--124

\bibitem{kose1984_En}
K.~Kose, H.~Hinata, Y.~Hashizume, E.~Futagawa, On a mathematical model of maneuvering motions of ships in low speeds, Journal of the Society of Naval Architects of Japan \textbf{1984}(155), 132 (1984)

\bibitem{Hwang1980}
W.~yuan Hwang.
\newblock Application of system identification to ship maneuvering (1980)

\bibitem{kijima1990manoeuvring}
K.~Kijima, T.~Katsuno, Y.~Nakiri, Y.~Furukawa, On the manoeuvring performance of a ship with theparameter of loading condition, Journal of the society of naval architects of Japan \textbf{1990}(168), 141 (1990)

\bibitem{Fujii_rudder_coef}
H.~FUJII, T.~TUDA, Experimental researches on rudder performance. (2), Journal of Zosen Kiokai \textbf{1961}(110), 31 (1961)

\bibitem{Ghadimi2013}
P.~Ghadimi, A.~Dashtimanesh, Y.~Faghfoor~Maghrebi, Initiating a mathematical model for prediction of 6-dof motion of planing crafts in regular waves, International Journal of Engineering Mathematics \textbf{2013} (2013)

\bibitem{Kerdraon2020}
P.~Kerdraon, B.~Horel, P.~Bot, A.~Letourneur, D.~{Le Touzé}, Development of a 6-dof dynamic velocity prediction program for offshore racing yachts, Ocean Engineering \textbf{212}, 107668 (2020)

\bibitem{Savitsky1964}
D.~Savitsky, {Hydrodynamic Design of Planing Hulls}, Marine Technology and SNAME News \textbf{1}(04), 71 (1964)

\bibitem{Troesch1992}
A.W. Troesch, {On the Hydrodynamics of Vertically Oscillating Planing Hulls}, Journal of Ship Research \textbf{36}(04), 317 (1992)

\bibitem{Hicks1995}
J.D. Hicks, A.W. Troesch, C.~Jiang, {Simulation and Nonlinear Dynamics Analysis of Planing Hulls}, Journal of Offshore Mechanics and Arctic Engineering \textbf{117}(1), 38 (1995)

\bibitem{Hamada2023}
S.~Hamada, Y.~Miyauchi, Y.~Akimoto, N.~Umeda, A.~Maki, {System identification of porpoising dynamics of high-speed planing craft using full scale trial data}, Ocean Engineering \textbf{270}, 113585 (2023)

\bibitem{Gertler1967}
M.~Gertler, G.R. Hagen, Standard equations of motion for submarine simulation, NSRDC report \textbf{2510} (1967)

\bibitem{Feldman1979}
J.~Feldman, Dtnsrdc revised standarrd submarine equations of motion.
\newblock Tech. rep., DAVID W TAYLOR NAVAL SHIP RESEARCH AND DEVELOPMENT CENTER BETHESDA MD. (1979)

\bibitem{Gertler1967PMM}
M.~Gertler, The dtmb planar-motion-mechanism system.
\newblock Tech. rep., David W Taylor Naval Ship Research and Development Center BETHESDA MD (1967)

\bibitem{maki2018UUV}
A.~Maki, T.~Tsutsumoto, Y.~Miyauchi, Fundamental research on the maneuverability of the underwater vehicle having thrust vectoring system, Journal of Marine Science and Technology \textbf{23}(3), 495 (2018)

\bibitem{Landweber1951}
L.~Landweber, J.~Johnson, Prediction of dynamic stability derivatives of an elongated body of revolution. revision.
\newblock Tech. rep., DAVID W TAYLOR NAVAL SHIP RESEARCH AND DEVELOPMENT CENTER BETHESDA MD (1951)

\bibitem{whicker1958}
L.F. Whicker, L.F. Fehlner, Free-stream characteristics of a family of low-aspect-ratio, all-movable control surfaces for application to ship design.
\newblock Tech. rep., David Taylor Model Basin Washington DC (1958)

\bibitem{dempsey1977}
E.M. Dempsey, Static stability characteristics of a systematic series of stern control surfaces on a body of revolution.
\newblock Tech. rep., David W Taylor Naval Ship Research and Development Center BETHESDA MD (1977)

\bibitem{Murakami1975En}
T.~Murakami, An investigation of the longitudinal motion of the slender-type body at fully submerged condition, Journal of the Society of Naval Architects of Japan \textbf{1975}(138), 256 (1975)

\bibitem{Murakami1978En}
T.~Murakami, An investigation of the longitudinal motion of the slender-type body at fully submerged condition, Journal of the Society of Naval Architects of Japan \textbf{1978}(143), 201 (1978)

\bibitem{Murakami2008En}
T.~Murakami, An investigation of longitudinal motion of the slender-type body at fully submerged condituin, Journal of the Japan Society of Naval Architects and Ocean Engineers \textbf{7}, 123 (2008)

\bibitem{Tokugawa1954En}
T.~Tokugawa, F.~Kito, Analysis of diving of a submarine, Journal of Zosen Kiokai \textbf{1954}(94), 69 (1954)

\bibitem{Carrica2013a}
P.M. Carrica, F.~Ismail, M.~Hyman, S.~Bhushan, F.~Stern, Turn and zigzag maneuvers of a surface combatant using a urans approach with dynamic overset grids, Journal of Marine Science and Technology \textbf{18}, 166 (2013)

\bibitem{Chase2013}
N.~Chase, P.M. Carrica, Submarine propeller computations and application to self-propulsion of darpa suboff, Ocean Engineering \textbf{60}, 68 (2013)

\bibitem{Motora1960En}
S.~Motora, On the measurement of added mass and added moment of inertia for ship motions (part 2. added mass abstract for the longitudinal motions), Journal of Zosen Kiokai \textbf{1960}(106), a59 (1960)

\bibitem{Motora1960_2En}
S.~Motora, On the measurement of added mass and added moment of inertia for ship motions (part 3. added mass for the transverse motions), Journal of Zosen Kiokai \textbf{1960}(106), a63 (1960)

\bibitem{Yoshimura2011_En}
Y.~Yoshimura, Y.~Masumoto, Hydrodynamic force database with medium high speed merchant ships including fishing vessels and investigation into a manoeuvring prediction method, Journal of the Japan Society of Naval Architects and Ocean Engineers \textbf{14}, 63 (2011)

\bibitem{Hasegawa1980}
K.~Hasegawa, On a performance criterion of autopilot navigation, Journal of the Kansai Society of Naval Architects \textbf{178}, 93 (1980)

\bibitem{Fujiwara1998_En}
T.~Fujiwara, M.~Ueno, T.~Nimura, Estimation of wind forces and moments acting on ships, Journal of the Society of Naval Architects of Japan \textbf{1998}(183), 77 (1998)

\bibitem{HUANG2020451}
Y.~Huang, L.~Chen, P.~Chen, R.R. Negenborn, P.~{van Gelder}, Ship collision avoidance methods: {State-of-the-art}, Safety Science \textbf{121}, 451 (2020)

\bibitem{Maki2014ASR_En}
A.~Maki, T.~Katayama, T.~Hasegawa, Tank test results of the surface ship (manoeuvring test results for an auxiliary vessel), Technical Research and Development Institute, Ministry of Defense technical report (7149-7172), 1 (2014)

\bibitem{Araki2012a}
M.~Araki, H.~Sadat-Hosseini, Y.~Sanada, K.~Tanimoto, N.~Umeda, F.~Stern, Estimating maneuvering coefficients using system identification methods with experimental, system-based, and cfd free-running trial data, Ocean Engineering \textbf{51}, 63 (2012)

\bibitem{Wang2020}
Z.~Wang, C.G. Soares, Z.~Zou, Optimal design of excitation signal for identification of nonlinear ship manoeuvring model, Ocean Engineering \textbf{196}, 106778 (2020)

\bibitem{Sutulo2014a}
S.~Sutulo, C.G. Soares, An algorithm for offline identification of ship manoeuvring mathematical models from free-running tests, Ocean Engineering \textbf{79}, 10 (2014)

\bibitem{Yoon2003}
H.K. Yoon, K.P. Rhee, Identification of hydrodynamic coefficients in ship maneuvering equations of motion by estimation-before-modeling technique, Ocean Engineering \textbf{30}, 2379 (2003)

\bibitem{Tsukada2014_En}
Y.~Tsukada, M.~Ueno, K.~Tanizawa, Y.~Kitagawa, H.~Miyazaki, R.~Suzuki, Development of an auxiliary thruster for free-running model ship tests, Journal of the Japan Society of Naval Architects and Ocean Engineers \textbf{20}, 59 (2014)

\bibitem{Ueno2014_En}
Y.T. Michio~Ueno, On rudder effectiveness and speed correction for free-running model ship tests, Conference proceedings, the Japan Society of Naval Architects and Ocean Engineers \textbf{18}, 235 (2014)

\bibitem{IEC62065}
I.E. Commission.
\newblock Iec 62065:2014, maritime navigation and radio communication equipment and systems - track control systems - operational and performance requirements, methods of testing and required test results (2014)

\bibitem{sawada_dissertation}
R.~Sawada, Research on control and sensing techniques of automatic berthing.
\newblock Ph.D. thesis (2023).
\newblock \doi{https://doi.org/10.18910/92962}

\bibitem{ITTC2002}
H.~Kobayashi, J.J. Blok, R.~Barr, Y.S. Kim, J.~Nowicki, {The Specialist Committee on Esso Osaka Final Report and Recommendations to the 23rd ITTC}, 23rd International Towing Tank Conference \textbf{II}, 581 (2002)

\bibitem{KOBAYASHI1994}
H.~Kobayashi, A.~Ishibashi, K.~Shimomkawa, Y.~Shimura, {A Study on Mathematical Model for the Maneuvering Motions of Twin-propeller Twin-rudder Ship : In Reference to the Maneuvering Motion from Ordinary Speed Range to Low Speed Range}, The Journal of Japan Institute of Navigation \textbf{91}(0), 263 (1994)

\bibitem{Yasukawa2003}
H.~Yasukawa, K.~Kose, {Simulation of Stopping Maneuver of a Tanker in Wind and Waves}.
\newblock in \emph{Transactions of the West-Japan Society of Naval Architects}, vol. 106 (2003), vol. 106, pp. 57--68

\bibitem{Ueno2001}
M.~Ueno, T.~Nimura, H.~Miyazaki, T.~Fujiwara, K.~Nonaka, H.~Yabuki, {Model Experiment and Sea Trial for Investigating Manoeuvrability of a Training Ship}, Journal of the Society of Naval Architects of Japan \textbf{2001}(189), 71 (2001)

\bibitem{Hasegawa1994}
K.~Hasegawa, T.~Fukutomi, {On Harbour Manoeuvring and Neural Control System for Berthing with Tug Operation}.
\newblock in \emph{Proc. of 3rd International Conference Manoeuvring and Control of Marine Craft (MCMC'94)} (1994), pp. pp.197--210

\bibitem{Yoshimura1978}
Y.~Yoshimura, K.~Nomoto, Modeling of manoeuvring behaviour of ships with a propeller idling, boosting and reversing, Journal of the Society of Naval Architects of Japan \textbf{1978}(144), 57 (1978)

\bibitem{Kitagawa2015}
Y.~Kitagawa, Y.~Tsukada, H.~Miyazaki, {A Study on Mathematical Models of Propeller and Rudder under Maneuvering with Propeller Reverse Rotation}, Conference Proceedings The Japan Society of Naval Architects and Ocean Engineers \textbf{20}, 117 (2015)

\end{thebibliography}

\end{document}